\documentclass[aps,prb,twocolumn,showpacs,superscriptaddress]{revtex4}
\usepackage{graphicx}
\usepackage{color}

\begin{document}

\title{Rationale for a Correlated Worldline Theory of Quantum Gravity}

\author{ P.C.E. Stamp}
\affiliation{Department of Physics and Astronomy, and Pacific
Institute of Theoretical Physics, University
of British Columbia, 6224 Agricultural Rd., Vancouver, B.C., Canada
V6T 1Z1}


\begin{abstract}
It is argued that gravity should cause a breakdown of quantum mechanics, at low energies, accessible to table-top experiments. It is then shown that one can formulate a theory of quantum gravity in which gravitational correlations exist between worldline or worldsheet paths, for the particle or field of interest. Using a generalized equivalence principle, one can give a unique form for the correlators, yielding a theory with no adjustable parameters.  A key feature of the theory is the "bunching" of quantum trajectories caused by the gravitational correlations - this is {\it not} a decoherence or a "collapse" mechanism. This bunching causes a breakdown of the superposition principle for large masses, with a very rapid crossover to classical behaviour at an energy scale which depends on the physical structure of the object. Formal details, and applications of the theory, are kept to a minimum in this paper; but we show how physical quantities can be calculated, and give a detailed discussion of the dynamics of a single particle.

\end{abstract}


\maketitle


\section{ Introduction}
 \label{sec:intro}


In 1957 Feynman remarked \cite{feynman57} that there were was an apparent conflict between Quantum Mechanics (QM) and General Relativity (GR), if one cared to extrapolate QM superpositions to the macroscopic scale; and various authors have since refined and extended the discussion of this conflict \cite{karolhazy66,hannay77,kibble79-81,kibble80,page79,unruh86,penrose84,penrose96}. There have been also been attempts to resolve it in theories wherein gravity modifies QM - most notably in a theory of Kibble et al. \cite{kibble79-81,kibble80}, and a set of arguments by Penrose \cite{penrose84,penrose96}, following on from results of Karolhazy \cite{karolhazy66}. There have also been several 'stochastic collapse' theories of QM which have attempted to involve gravity \cite{diosi90,bassi-grav,pearle-grav}, although none of these is fully relativistic.

All these discussions assume that QM, rather than GR, will break down at the energy scales accessible to lab experiments. Given that GR works extremely well at low energies, with no sign in any experiments or observations of departures from standard torsionless Einstein theory (and none of the many attempts \cite{ferreira12,capoz11,sotiriou10,odintsov} made to modify low-energy GR have yet yielded a viable competitor), this seems at least a reasonable point of view.

However QM also works incredibly well in this regime, and has also never failed laboratory tests. Since both GR and QM work so well, most physicists would prefer to ignore the apparent low-energy conflict between them, or conclude that it does not actually exist. We emphasize that there is an internally consistent low-energy effective theory of quantum gravity describing quantum fluctuations around classical spacetime \cite{donoghue94,barvinsky} (although it has hardly been tested experimentally).

On the other hand at very high energies, approaching the Planck energy $E_p = M_pc^2 \sim  1.22 \times 10^{28}$eV, it is generally believed that GR will fail, given the well-known problems of perturbative quantum gravity \cite{GtH74,goroff}. We shall have little to say here about physics at such high energies (at $\sim 10^{17}$ times the energy of current experiments - a much greater gap than that separating us from the pre-Babylonian world), nor about new physics that might exist between current energies and the Planck scale.

There is also a growing belief that QM should also eventually fail, this time at macroscopic scales - this is because of the well-known paradoxes associated with the application of the superposition principle at these scales. However this belief is still rejected by most physicists, who prefer to modify our ideas about the macroscopic world to agree with QM, and to assume that GR must break down when it conflicts with QM.

What I will instead argue is that the low-energy conflict between GR and QM is real, and that the correct way to resolve it is to modify QM. I will then show that a new theory can be written in terms of a specific kind of correlation between paths in any QM path integral, with the correlations caused by gravity; these ideas were first presented in a sketchy way in ref. \cite{stamp12}. We see how one can fix the form of these correlations, leading to a theory with no adjustable parameters, which makes quantitative predictions for low-energy phenomena. This theory is essentially a low-energy theory of quantum gravity, in which conventional QM is violated. However these violations only turn out to be appreciable on the macroscopic scale, in regimes where they have yet to be tested.

The main focus of this paper is on the physical principles and arguments leading to this "Correlated Worldline" (CWL) theory; an attempt is also made to give some intuition for how it works. The readership is assumed to be fairly broad (gravity, condensed matter, atomic physics), so technical details are kept brief, and the mathematics fairly simple (a much more complete discussion of the theory appears elsewhere \cite{GR-Psi-L1,GR-Psi-1}). Nevertheless some of the formalism is described, to show how CWL theory differs from standard low-energy quantum gravity, and how it can lead to experimental predictions. This is an interesting time for laboratory work on this topic - several papers have proposed experiments \cite{bouw03,aspel-exp,bouw08}, and there is some optimism that they may be able to test alternative theories. The detailed experimental ramifications of the CWL theory are discussed in other papers \cite{GR-Psi-L1,queisser14}.

The plan of the paper is as follows: section 2 reviews the arguments indicating a fundamental conflict between GR and QM at low energies, as well as some of the internal problems of each theory; and section 3 discusses how one might resolve this conflict - in these sections the previous literature is also reviewed. Sections 4 and 5 describe the CWL theory, and are central to the paper. Section 4 describes the key arguments, along with a set of postulates which formulate these more precisely. Section 5 then describes the basic features of CWL theory, giving enough details so one can see how physical quantities are derived, and including a very simple example to show how things work out quantitatively. Section 6 summarizes the results.


\section{The Conflict between Gravity and Quantum Mechanics}
 \label{sec:GR-QM}


As already noted, the question of whether or not GR and QM are in conflict at low energies is controversial. We therefore begin by reviewing this important question, and make the case that the conflict is real.

\subsection{Internal Problems within General Relativity and Quantum Theory}
 \label{sec:GR-QM1}

It is always worth recalling that both GR and QM have their own individual problems. Key amongst these are:

\vspace{2mm}

(i) {\it Problems in classical General Relativity}: Classical GR is, mathematically, a perfectly consistent theory - but the status of both the spacetime metric $g^{\mu \nu}(x)$ and the affine connection $\Gamma_{\mu \nu}^{\alpha}(x)$ have been debated since it was first formulated \cite{einstein16,weyl18,cartan22,fock59,synge60} (more recent perspectives are summarized in refs. \cite{norton99,rovelli,brown,lehmkuhl14}). Einstein's early view \cite{einstein16}, expressed in his famous 'hole argument', was that spacetime has no independent existence, and is instead defined by 'spacetime coincidences' involving matter. His later remarks sometimes contradicted each other; but his initial view had to be modified, if for no other reason than the existence of Weyl degrees of freedom, independent of matter. This leads naturally to the idea that spacetime should be viewed as a field. However it is like no other field, coupling to all of them, and to itself, in the same way (the 'weak equivalence principle'), and also providing the underlying background geometry, and underlying causal structure, upon which all other fields are supposed to live. Moreover, although the spacetime metric measures the local energy-momentum of all other fields, its own energy-momentum cannot be localized \cite{gravLoc}.

Mathematically, the connection and the metric are quite distinct, and one can argue that they should also be considered as ontologically distinct - Einstein apparently thought this at least part of the time \cite{lehmkuhl14} - even though in standard GR they are united by the Levi-Civita relation. One can of course extend GR to include a torsion field $2S_{\mu \nu}^{\alpha} = \Gamma_{\mu \nu}^{\alpha}(x) - \Gamma_{\nu \mu}^{\alpha}(x)$ and 'non-metricity' field $Q_{\alpha \mu \nu}(x) = \nabla_{\alpha} g_{\mu \nu} (x)$. Purely torsional theories have many attractive features, and a long history \cite{torsion}, but their consequences for experiments or astronomical observations are not currently believed to be important \cite{tegmark07}.

Most authors assume that GR must ultimately be internally inconsistent, because of the inevitability \cite{penroseH} of singular solutions to the field equations (which some Quantum Field Theory (QFT) might somehow cure). Torsion has also been suggested as a way to avoid singularities \cite{tor-sing}. On the other hand, one can also argue that singularities are actually necessary for the stability of the theory \cite{myers}. The current evidence - both in the many weak-field tests \cite{will14,will}, and in the enormous variety of predicted strong field phenomena, which have revolutionized astronomy, cosmology, and astrophysics - shows overwhelmingly that classical GR is valid up to very high energies (still $\ll E_p$). On the other hand, any attempt to quantize it leads to huge problems, as we discuss below.

\vspace{2mm}

(ii) {\it Problems in Quantum Mechanics}: The internal problems of QM are much more severe than those of GR, and have been discussed repeatedly \cite{laloe,bell}. In QM or in QFT a system in a pure state is described, according to the "state-vector as physics" viewpoint, by a state vector $|\psi(t)\rangle$, along with the Hamiltonian (or Lagrangian) determining its dynamics. However any claim that $|\psi(t)\rangle$ then represents something physically real is untenable - changes in $\langle \{ {\bf r}_j \}|\psi(t)\rangle$ can happen non-locally (as in, eg., the EPR paradox), violating special relativity for a physically real object. On the other hand a "state vector as knowledge" view, which treats $|\psi(t)\rangle$ solely as a representation of our knowledge of the system, loses any connection to physical reality, and/or treats it anthropocentrically. Moreover, this interpretation is ambiguous - different observers can have different information about a system.

Many authors then take the view that the state-vector is more trouble than it is worth, and that one should work only with the density matrix; others simply treat QM as a theory of correlations between measurements. The fall-back position is then to only talk about measurements, and operators as means of defining measurements - this "measurement calculus" view is probably the most popular. Such approaches simply accept from the start that QM can only discuss probabilities, that the state vector is just an artefact of the theory, and that all we need, FAPP (ie, "For All Practical Purposes"), are these probabilities and correlations. "Physical Reality" is then typically treated as a needless abstraction.

These interpretations run into severe paradoxes (eg., the 'Schrodinger's Cat', or 'Wigner's friend' paradoxes) when dealing with macroscopic superpositions of states - it is not clear what $|\psi(t)\rangle$ is then referring to, nor how one is supposed to deal with the definite classical states in which macroscopic objects appear to exist. For a long time the Copenhagen interpretation (in which the world was divided in a rather mysterious way into classical and quantum worlds) was supposed to take care of this - but this distinction becomes ever less obvious as QM is pushed towards the macroscopic scale.

QM is tested daily at the microscopic scale, in a huge variety of circumstances. So far all Bell inequality tests \cite{belltest} on widely-separated systems have only involved microscopic entanglement. Tests of QM at the macroscopic scale are far less frequent. The idea of macroscopic quantum  phenomena and macroscopic wave-functions goes back to London, in his early discussions \cite{london-sfl} of superfluids and superconductors; London was well aware of the fundamental implications for QM \cite{londonBauer}. Some attempts have been made to characterize just how macroscopic a given quantum state may be \cite{AJL80,Lgarg,cirac,whaley}; however different authors give very different answers \cite{whaley}. In any case, experiments showing quantum behaviour at levels well above the atomic scale have been done on several systems, notably superconductors \cite{scon} and spin arrays \cite{SMM}. Tests of the "Leggett-Garg" inequalities (a temporal analogue, for a single quantum system, of Bell's inequalities) have also shown no breakdown of QM \cite{nori}. Thus, insofar as large-scale QM superpositions have yet been tested, QM has been vindicated.

However one thing that has not been tested at all is the validity of QM when superpositions of states involving large mass displacements are involved. This is of course where the conflict with our own experience of the world becomes most severe - it is also where gravitational effects might be expected to arise.

\subsection{Quantizing Gravity}
 \label{sec:quant-GR}

The problem of quantizing gravity encounters many technical difficulties - here we attempt to disentangle these from questions of principle. Both have a bearing on the question at issue, viz., on the compatibility of the two theories.

Efforts to quantize Einstein gravity go back to the 1930's \cite{rosen30}; the field has repeatedly been reviewed \cite{rovelli,brill70,ashtekar74,alvarez89,carlip01,kiefer06,woodard09}.
By "conventional" Quantum Gravity we will mean here a theory that starts from a generating functional
\begin{equation}
{\cal Z}[J_{\mu \nu}] \;=\; \int {\cal D}\tilde{\mathfrak{g}}^{\mu \nu}(x) \int {\cal D} \phi \; \Delta[\tilde{\mathfrak{g}}^{\mu \nu}(x)] \exp [{i \over \hbar} S]
 \label{Z-gr}
\end{equation}
where $\Delta[\tilde{\mathfrak{g}}^{\mu \nu}(x)]$ is a Fadeev-Popov determinant, $\tilde{\mathfrak{g}}^{\mu \nu}(x) = g^{1/2}(x) g^{\mu \nu}(x)$ is the metric density tensor, and the action terms are
\begin{equation}
S = {\cal S}_G + {\cal S}_{\phi} \;+\; \int d^4x \; {1 \over \lambda} J_{\mu \nu}(x) \; \tilde{\mathfrak{g}}^{\mu \nu}(x)
 \label{S-gr}
\end{equation}
(here $S_{\phi}$ describes all matter fields, $S_G$ the gravitational degrees of freedom, $\lambda^2 = 16 \pi G$, and $J_{\mu \nu}$ is an external source). As always with path integrals, there is a normalization factor which we do not write explicitly (however we will write it when we come to the CWL theory).

One can if desired extend this to torsional and non-metric terms in the action - there is then a functional integral over the connection, and one may introduce various Lagrange multiplier fields to enforce constraints.

All attempts to quantize conventional gravity defined in this way, with $\tilde{\mathfrak{g}}^{\mu \nu}(x)$ taken simply as another field, run into well-known difficulties \cite{GtH74,goroff,carlip01,woodard09,wald84}. Because of quantum fluctuations of the metric density $\tilde{\mathfrak{g}}^{\mu \nu}(x)$, including topology-changing contributions, one loses the causal structure of the spacetime geometry in the path integration over $\tilde{\mathfrak{g}}^{\mu \nu}(x)$; one must also deal with a conformal instability, and so on - the functional integral is simply not well-defined. Since the causal structure of the spacetime is required for any QFT living in it, relativistic QFT then becomes meaningless (note, in this connection, that although the Einstein GR action is the only reasonable candidate for a theory built from spin-2 gravitons \cite{deser70,wald86,feynman63,weinberg64}, the causal structure changes fundamentally once gravitational fluctuations are allowed to interact \cite{penrose76}).

If the generating functional in (\ref{Z-gr}) is viewed only as a means of generating perturbative expansions, the quadratically divergent high-energy contributions render the theory non-renormalizable, with or without matter \cite{GtH74,goroff}. One can speculate that there may be some non-perturbatively renormalizable "asymptotically safe" version of quantum gravity, wherein divergences are somehow self-consistently suppressed. Such ideas are old \cite{kriplo66}, and were revived both in Weinberg's 'asymptotically safe' scenario \cite{weinberg-asym}, and in more recent calculations of Davidson and Wilczek \cite{wilczek06}, Toms \cite{toms-asymp}, and others \cite{donoghue12}. One can try to check such ideas using sophisticated diagrammatic approximations \cite{bern}, or using a variety of non-perturbative methods, often involving numerical work \cite{loll,eichhorn,reuter}.  Whether or not any of these scenarios can work, experimental or observational tests are extremely difficult - the energies involved are enormous.

A more common tactic has been to simply change the rules of the game: one regards GR, with or without matter, as a low-energy limit of something which is quite different at high energies. String theory and modern loop gravity fall into this category. One can then look for weak testable departures from classical GR at low energies, caused by inaccessible high-energy physics - this has led to a vast effort to devise low-energy "extended gravity" theories \cite{ferreira12,capoz11,sotiriou10,odintsov}, more recently motivated by cosmological observations.

One thing that has been learnt from such efforts is quite how hard it is to change GR without finding either instabilities, or acausal behaviour, or a breakdown of unitarity caused by ghost modes (a nice example being "massive gravity" theories \cite{hinter12}, where a constraint removes the Boulware-Deser ghost but then makes the theory acausal \cite{deser13}). Related lines of thought look at  "induced gravity" \cite{adler76}, or more general "emergent gravity" \cite{carlip13} scenarios; and there is a long tradition, recently revived, of looking for extended versions of classical GR embedded in higher dimensional spaces \cite{embeddedGR,lovelock}.

Alternatively, one can argue that quite new physics will intervene long before we get to Planck energies - the existence of dark matter is often taken as a hint that this may be the case, as are the continuing attempts to understand the value of the Higgs mass.

Whatever one might think of any of these approaches - and there is a huge variety here - two really crucial points clearly emerge, viz.,

\vspace{1mm}

(i) There is a well-defined low-energy theory of quantum gravity. If one assumes the validity of Einsteinian GR in the classical regime, then this is the perturbative structure formulated by Barvinsky, Vilkovisky et al. \cite{barvinsky}, and by Donoghue \cite{donoghue94}. One can in principle generalize it to include weak departures from Einsteinnian GR ("extended gravity" theories). This perturbative apparatus is neither simple nor without controversy \cite{shapiro-effG}; but it is certainly the best we have right now.

\vspace{1mm}

(ii) All these approaches - low-energy effective theories or more ambitious high-energy theories - assume that {\it quantum theory is universally valid}. Thus they are all competitors, in any experimental test, to the kind of theory we will discuss below, in which QM is violated.

\subsection{Problems quantizing Gravity in the Low-energy regime}
 \label{sec:lowE-QGr}

Orthodox Quantum Gravity assumes that GR (or some extended GR) is fine if viewed as a low-energy effective theory \cite{donoghue94,barvinsky} resulting from some high-energy quantum theory of strings, or loops, or something else. This view implies there will be no experimental conflict between QM and either low-energy quantized gravity, or with classical GR itself (since quantum corrections to classical GR are extremely small at low energies).

To see that this view may not be correct, we consider two thought experiments that clearly show there is a conflict between GR and QM at low energies.

\subsubsection{2-path interference with a mass $M$}
 \label{sec:thoughtExp}

\begin{figure}
\vspace{-0.6cm}
\includegraphics[width=\columnwidth]{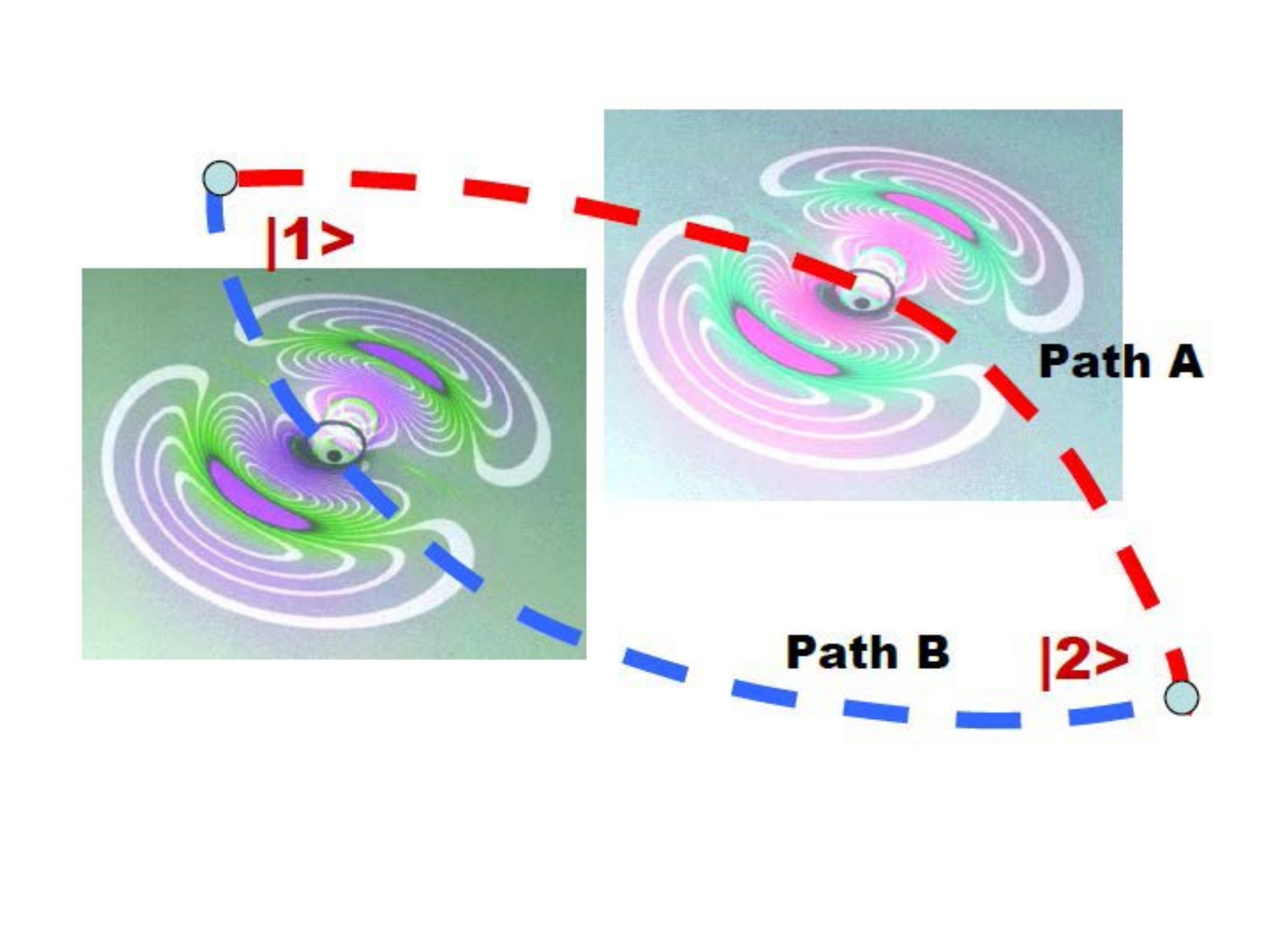}
 \vspace{-1.6cm}
\caption{The gravitational "2-path" experiment: an object of mass $m$ can follow one of two QM paths. Each path carries with it its own distortion of the spacetime metric, shown in caricature.  }
 \label{fig:NJP-F1}
\end{figure}

We set up a geometry in which a massive object is forced into a superposition of 2 different paths (Fig. \ref{fig:NJP-F1}). In conventional state vector language this means we are dealing with the superposition
\begin{equation}
|\Psi \rangle \;=\; a_1 |\Phi_1; \tilde{\mathfrak{g}}_{(1)}^{\mu \nu}(x) \rangle + a_1 |\Phi_2; \tilde{\mathfrak{g}}_{(2)}^{\mu \nu}(x) \rangle
 \label{GRsup}
\end{equation}
or the associated density matrix $\rho_{kk'}$, where $k,k' = 1,2$. In this superposition, those gravitational degrees of freedom that are tied to matter in the gravitational part $|\tilde{\mathfrak{g}}^{\mu \nu}(x) \rangle$ of the state vector are then completely entangled with $|\Phi\rangle$. Note that we cannot avoid incorporating the metric into the state superposition - as has been remarked on many occasions \cite{feynman57,hannay77,kibble79-81,kibble80,page79,unruh86}, once we tie the metric to matter, we must quantize the metric field.


We now immediately find a paradox. The superposition of the 2 spacetime metrics, assuming we can define it at all, leads to violation of energy-momentum conservation \cite{unruh86}, and/or   superluminal signal propagation \cite{hannay77}, depending on whether one allows this wave-function to collapse during a measurement. When this {\it gedankenexperiment} was first proposed \cite{feynman57}, it could perhaps forgivably be ignored - the then prevailing mythology forbade such macroscopic superpositions. Current evidence for macroscopic coherence effects in superfluid and magnetic systems \cite{scon,SMM} now makes such a view untenable.

Actually the problem is worse than these remarks indicate. When we write the gravitational state vector as $|\tilde{\mathfrak{g}}^{\mu \nu}(x) \rangle$, it appears in the form of a quantum field over spacetime coordinates $x^{\mu}$; but the relationship between different coordinates $x$ and $x'$  is not defined until we have already fixed the metric density $\tilde{\mathfrak{g}}^{\mu \nu}(x)$. If we take two different spacetimes, with two different metrics $\tilde{\mathfrak{g}}^{\mu \nu}(x_1)$ and $\tilde{\mathfrak{g}}^{\mu \nu}(x_2)$, and two different causal structures, then no relationship exists at all between the coordinates $x_1$ and $x_2$. Thus one can ask what it even means to superpose two different spacetimes - as Penrose has put it \cite{penrose96}, how does one even map between the two different manifolds on which these states apparently reside?

To see we are dealing here with a basic conflict between QM and GR in the low-energy regime (where GR is supposed to be accurate), consider the simple non-relativistic wave-function superposition
\begin{equation}
\Phi({\bf r},t) \equiv \langle {\bf r}|\Phi(t) \rangle \;=\; a_1 \Phi_1({\bf r},t) + a_2 \Phi({\bf r},t)
 \label{Phisup}
\end{equation}
What now is meant by the coordinates ${\bf r}, t$? If we treat spacetime as a background 'substrate' for the wave-function, completely unaffected by the presence of the object, then we recover standard QM. But this assumption is untrue, and the coordinates ${\bf r_1}, t_1$ and ${\bf r_2}, t_2$ involved in the two branches of the superposition cannot be identified with each other. So how do we then even {\it define} the relationship between the two components of the wave-function? We can no longer write the overlap between $\Phi_1({\bf r_1}, t_1)$ and $\Phi_2({\bf r_2}, t_2)$ as $\langle \Phi_1 | \Phi_2 \rangle$, because this presupposes that we can identify ${\bf r}_1$ with ${\bf r}_2$, and $t_1$ with $t_2$. Thus even the basic QM idea of interference loses all meaning, if we take GR seriously at low energies.

One can try to argue that this problem is academic because the effects are small and/or currently unmeasurable. This argument simply ignores a basic problem of principle; I will argue below that it is also incorrect.

\subsubsection{EPR experiment with 2 entangled Masses}
 \label{sec:thoughtExp2}

\begin{figure}
\vspace{0.2cm}
\includegraphics[width=\columnwidth]{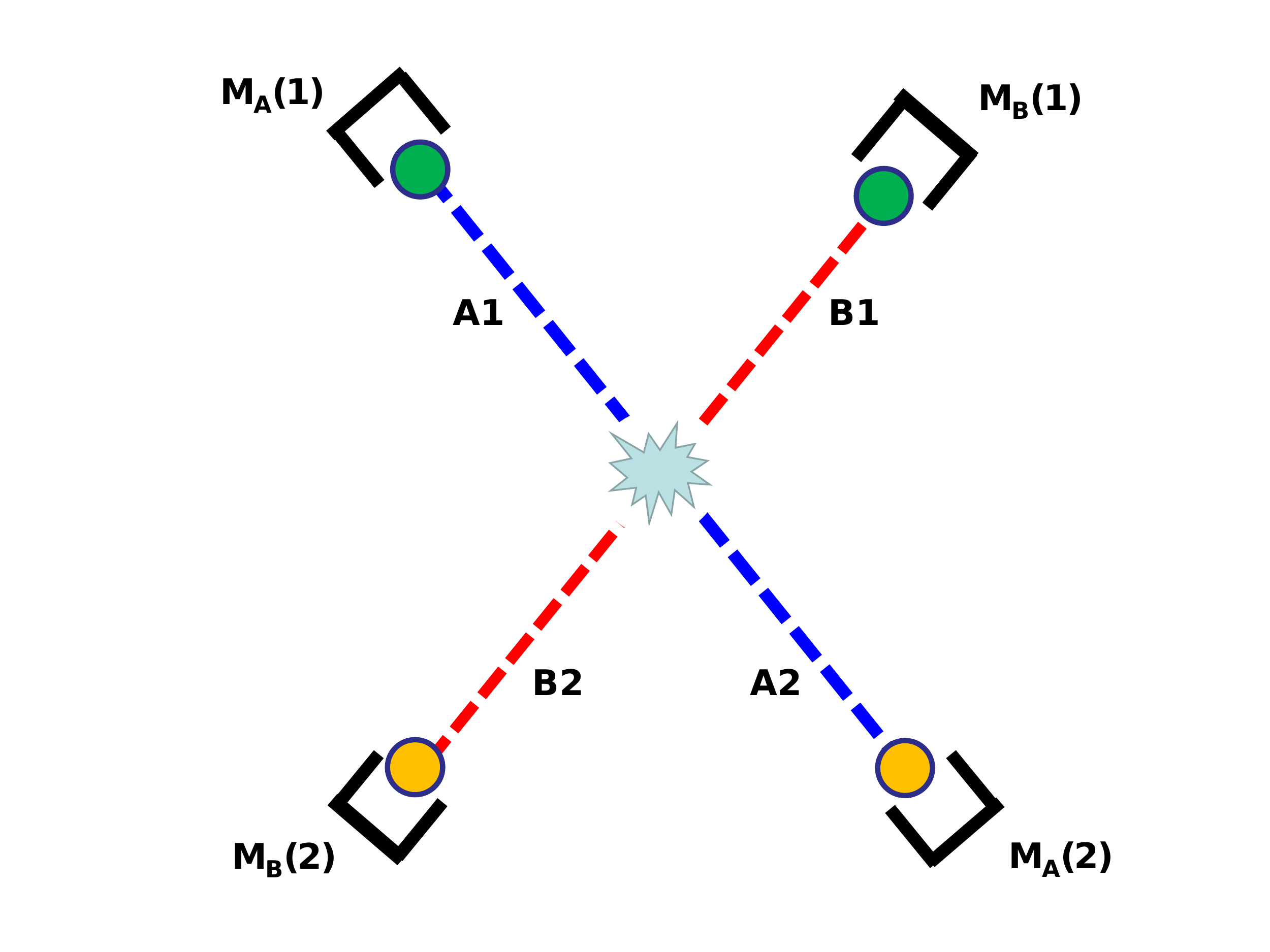}
 \vspace{-0.6cm}
\caption{An "Einstein-Podolsky-Rosen" thought experiment in which 2 masses, labeled "1" and "2", separate from the origin in a superposition of states having equal and opposite momenta. They are detected by 4 measuring systems. If mass 1 is observed by the measurement apparatus $M_A(1)$, mass 2 will be necessarily seen at $M_A(2)$; conversely if mass 1 is seen by $M_B(1)$, then mass 2 will necessarily be seen by $M_B(2)$. No influence or information can propagate between the 4 measuring systems in the time it takes them to make their measurements.   }
 \label{fig:NJP-F2}
\end{figure}

The problems just discussed become even more acute when dealing with an Einstein-Podolsky-Rosen (EPR) pair of masses - apparently this point has not been previously discussed, although it is rather obvious, so I just sketch it here. Imagine a pair of masses in a set-up similar to that of EPR \cite{EPR}, and depicted in Fig. \ref{fig:NJP-F2}. We write the initial pair state of the masses as
\begin{equation}
|\Psi(1,2)\rangle = {1 \over \sqrt{2}} [ |\psi_A(1) \psi_B(1)\rangle + |\psi_A(2) \psi_B(2)\rangle ]
 \label{EPR-M}
\end{equation}
so that they move in opposite directions, such that if system A arrives at detector $A_1$, then system B arrives at detector $B_1$, and likewise with the other 2 detectors. We now suppose that the 4 detector systems are very far apart, so that no signal can pass between them while measurements are being made at any one of them.


From this point on the argument parallels previous discussions \cite{hannay77,kibble80,unruh86}; thus, if we allow $|\Psi\rangle$ to collapse during a measurement, we get superluminal changes in both $|\Psi\rangle$ and the spacetime metric, so that the energy-momentum tensor $T^{\mu\nu}(x)$ associated with $\tilde{\mathfrak{g}}^{\mu \nu}(x)$ is not conserved, and so on. The new element that the EPR set-up brings here is the very wide separation of the masses (including those involved in $M_A$ and $M_B$), so that no possibility of causal connection exists. Unless we insist on treating everything quantum-mechanically (including the measuring systems), we have a very severe paradox.

\vspace{1mm}

Other arguments have made the same basic point - in particular, Unruh \cite{unruh86} gave an interesting discussion in which a neutron star acted as an efficient spacetime measuring device. All these arguments indicate that we have a fundamental conflict between GR and ordinary QM, at laboratory energies, where both theories are supposed to work. They also imply that low-energy effective theories of quantum gravity cannot be completely correct. However, they do not tell us how to resolve the conflict.

In what immediately follows I will describe some of the attempts that have been made by previous writers to resolve this conflict. Much later in the paper, in section \ref{sec:fin}, I will discuss how the CWL theory described in this paper resolves them.

\section{Resolving the Conflict: What can we change?}
 \label{sec:conflict}

We now look for a resolution of this {\it impasse}, by seeking a modification of low-energy QM. We begin by reviewing what has already been tried, and then ask what kind of a theory we really want.

\subsection{Previous work}
 \label{sec:mod-Sch}

There have been many attempts to modify QM over the years; however most have been trying to cure the measurement problem in non-relativistic QM, making no mention of gravity. Those that have incorporated gravity - a much smaller group of theories - all share one feature: they incorporate the state vector $|\psi\rangle$ in the theory from the beginning, and then try to modify its dynamics. They fall into 3 groups:

\subsubsection{Uncertainty Principle arguments}
 \label{sec:uncP}

The first use of QM uncertainty principle arguments involving full-scale GR gravity was apparently by Karolhazy \cite{karolhazy66}; there are now many analyses in the literature \cite{hu08,clock}. The coupling between the stress-energy tensor and the spacetime metric means that uncertainties in the position, energy, momentum, and time of any matter immediately translate into uncertainties in $g^{\mu \nu}(x)$ and $R^{\mu}_{\nu \alpha \beta}(x)$, whose effect clearly increases with the mass of the matter. More recently Penrose, starting from the superposition (\ref{GRsup}), argued \cite{penrose96} that in a certain Newtonian limit (where $c \rightarrow \infty$), for which the superposition is described by (\ref{Phisup}), the ambiguity in elapsed proper time between the two components of the state vector should translate into a time uncertainty $\Delta t_{12}^g$ (here the "$12$" subscript indicates the components $1$ and $2$ of the wave function).

In a further step, Penrose argued that this time uncertainty could be equated to a {\it decoherence time} $\tau_g^{\varphi}$.  Writing $\tau_g^{\varphi} = \hbar /\Delta E^g_{12}$, he argued that for the superposition (\ref{Phisup}), one should have
\begin{equation}
\Delta E^g_{12} \;=\;  4 \pi G \int d^3r \int d^3r' {\Delta \rho_{12}({\bf r})\Delta \rho_{12}({\bf r'}) \over |{\bf r} - {\bf r'}|} \;\;\;
 \label{penrose}
\end{equation}
where $\Delta \rho_{12}({\bf r}) = \rho_1({\bf r}) - \rho_2({\bf r})$, and the $\rho_j$, with $j=1,2$, are the mass density distributions associated with each component of the wave function. Based on this idea, proposals \cite{bouw03,aspel-exp} for an experimental search for such decoherence were made.

The result (\ref{penrose}) of Penrose is essentially an uncertainty estimate, and thus in principle no different in spirit from the original arguments of Karolhazy \cite{karolhazy66} (who did however base his analysis in GR rather than any Newtonian limit). However, it is a big step to go from a $\Delta t$ derived from a time-energy uncertainty and call this a decoherence time - note that no decoherence is implied by the standard time-energy uncertainty relation in QM. The basic idea of Penrose is that we are dealing here with an {\it intrinsic decoherence} in Nature, equivalent to a breakdown of QM. The idea of intrinsic decoherence has been discussed by other authors \cite{milburn,GtH99,stamp06}, but no full scale theory has yet been proposed for it.

There are a number of important problems associated with eqtn. (\ref{penrose}). In particular:

(i) The Newtonian results cannot be extrapolated to strong fields - as Penrose \cite{penrose96} emphasized, this would involve a mapping between 2 distinct spacetime manifolds, and no such unique mapping exists.

(ii) More important, the result (\ref{penrose}) is undefined. For a single point particle it diverges, and otherwise depends on the form adopted for the $\rho_j({\bf r})$, ie., on whatever 'size' is adopted for the 'elementary' particles in the system concerned - this lengthscale is of course dependent on whatever UV cutoff we adopt. This is obvious in the widely varying estimates for $\Delta E_{ij}^g$ made by the Bouwmeester group \cite{bouw08} (see also Adler \cite{adler07}). Thus it is not clear what the expressions in (\ref{penrose}) and (\ref{diosi}) really mean, and they certainly do not provide unambiguous testable predictions. We return to this topic in section \ref{sec:CWL-GF} below.

We see that, as in the last section, uncertainty principle arguments indicate that we have a problem to solve, but they do not tell us how to solve it. Nevertheless, this work, in the form of eqn. (\ref{penrose}), has provided a clear target for experimentalists \cite{bouw03,aspel-exp,bouw08}. The most promising experimental designs for this work seem to be optomechanical, which are currently able to look at much larger masses than those in the very elegant multi-slit diffraction experiments pioneered in Vienna \cite{arndt}.

\subsubsection{Stochastic Collapse approaches}
 \label{sec:stochastic}

In a quite different approach, now involving a breakdown of QM, one modifies the Schrodinger equation by adding stochastic terms to it, to produce an equation of motion for $|\psi\rangle$ of form
\begin{equation}
(\hat{\cal H} - i \hbar \partial_t) \psi({\bf r},t)  =  \xi(\langle \psi| F(\hat{M}_j(t))|\psi \rangle, t)
 \label{Sch-var}
\end{equation}
where the argument of the 'noise' term $\xi(\langle F \rangle)$ involves a set of operators $\hat{M}_j(t)$ acting in the Hilbert space of the system. Work of this kind \cite{bassi07,bassi12}, motivated by the quantum measurement problem, almost always deals with non-relativistic QM, and assumes the usual QM structure of state vectors, Hilbert space, and projective quantum measurements. The noise term may depend on $|\psi \rangle$, making (\ref{Sch-var}) non-linear. Note that in contrast to any uncertainty principle arguments, this approach definitely leads to decoherence, and "wave-function collapse", caused precisely by the stochastic noise field.

Diosi \cite{diosi90,diosiR}, Ghirardi et al. \cite{ghirardi} and Pearle \cite{pearle} have discussed the possibility of a gravitational origin for this noise. Assuming a non-relativistic Newtonian framework, Diosi gets an energy "smearing"
\begin{equation}
\Delta E^g_{12} \;=\; 4 \pi G \int d^3r \int d^3r' {\rho_{1}({\bf r})\rho_{2}({\bf r'}) \over |{\bf r} - {\bf r'}|}
 \label{diosi}
\end{equation}
which can be translated into a decoherence time. Equations (\ref{penrose}) and (\ref{diosi}) are very similar; however the underlying assumptions and motivations are rather different. The wave-function collapse theories are attempting to deal with one specific problem, viz., the measurement problem, and give a definite mechanism leading to decoherence (although typically there are undetermined parameters in the theory). Penrose's arguments, although confined to the Newtonian limit, are an attempt to discern just what a much more general theory might look like, starting from general principles. Note that expression (\ref{diosi}) also has the same inherent problems as (\ref{penrose}), viz., it is purely Newtonian, and is mathematically ill-defined.

Attempts have been made to update expressions like (\ref{diosi}). Strong criticism of Diosi's original ideas \cite{bassi-grav} led to the more recent formulations of Diosi \cite{diosiR} and Pearle \cite{pearle-grav}. However, all stochastic theories of "gravitational decoherence" are subject to (at least) two fundamental problems, viz.:

(i) They are not relativistic. A fully relativistic theory needs to satisfy far more than just Lorentz invariance: it has to be a genuine QFT, which avoids the problems besetting almost all potential QFTs, including ghosts, acausality, violation of one or other conservation laws or symmetries known to be true, etc. And this is before one even addresses the requirement of general covariance.

(ii) They are not {\it generally applicable}. They do provide a new physical mechanism, involving non-relativistic stochastic fields. But they address a single problem (the measurement problem) in isolation; no general reason is given for introducing these fields, and their effect on other physical phenomena is often barely discussed. It is hard to see how extra fields introduced in this way are not going to conflict with other parts of physics; they should at any rate have testable effects on a huge variety of other physical phenomena.

\subsubsection{Non-linear Schrodinger equations}
 \label{sec:nonL}

As an alternative to stochastic modifications of the Schrodinger equation, one can keep it deterministic but make it non-linear. Early attempts \cite{bialynicki} to write a concrete non-linear generalization of the non-relativistic Schrodinger equation, in the form
\begin{equation}
(\hat{\cal H} - i \hbar \partial_t) \psi({\bf r},t) \;=\; f(\psi({\bf r},t), \psi^{\dagger}({\bf r},t))
 \label{NLSE}
\end{equation}
found this was not so easy; the rescaling of the wave-function during measurements required a specific logarithmic form for the potential, and it is hard to make theories of this kind consistent if one also assumes conventional ideas about quantum measurements. Nevertheless such attempts continued until Weinberg's work \cite{weinberg89}, which had two great virtues: it encapsulated many rather general features that such non-linear generalizations of Schrodinger's equation should possess, and at the same time produced a specific theory which was experimentally testable (and indeed it was falsified within a year \cite{weinbergT}). The problems afflicting any such non-linear QM theory were emphasized at the same time by Gisin, Polchinski, and others \cite{polchinski90}.

A different line of argument was taken by Kibble et al. \cite{kibble79-81,kibble80}; rather than adding non-linear terms to the Schrodinger equation, they introduced a specific mechanism, in which the non-linear structure of GR enters directly into the theory. This idea has very attractive features - it no longer simply involves {\it ad hoc} addition of terms to an existing structure like the Schrodinger equation, but instead attempts to change the structure of relativistic QFT in a way reflecting basic features of GR. Thus the ideas of Kibble et al. were very much in tune with the whole enterprise of quantum gravity \cite{carlip01,woodard09,kiefer06}; however, unlike conventional quantum gravity, Kibble et al. no longer assumed the superposition principle to be valid (compare equations (\ref{Z-gr}), (\ref{S-gr}) above).

To implement this idea turned out to be extremely difficult, and the result, written in terms of the total wave-function, was not in the form of a definite theory - this is also true of a few further attempts in this direction \cite{elze07}. One of the main reasons for the problems and inconsistencies encountered by Kibble et al. was that they kept too much of the conventional structure of QM, in the form of wave-functions, measurements, operators, etc. (and this handicap was emphasized by Kibble \cite{kibble79-81}). This then prevented them from making a real break from QM. Otherwise this is the only attempt so far which has genuinely tried to derive a new theory from first principles, and to incorporate fully relativistic gravity into the theory, along with the severe non-linearity coming from GR.

\subsection{Desiderata for a New Theory}
 \label{sec:new}

What lessons can we draw from all this? One cannot help feeling that many of the approaches just discussed are both too radical and not radical enough. Attempts to modify QM which still keep most of its baggage (the Schrodinger equation, the state vector, Hilbert space, measurements, and so on) are simply grafting extra terms onto an already problematic structure: they are not radical enough. Approaches introducing models which cannot encompass QFT or GR then disregard the most successful parts of modern physics: this is certainly too radical. Conventional quantum gravity simply assumes QM or QFT to be universally valid, thereby ignoring all the arguments given above for their inadequacies: this is not radical enough.

Let us therefore demand that:

(i) Paradoxically, because both QM and GR are so overwhelmingly successful, we cannot just reject one in favour of the other. Nor can we just drop inconvenient parts: the unified interlocking nature of physical theory will not tolerate small changes inconsistent with the whole. Any replacement theory has to involve new mechanisms, motivated by general physical considerations.

(ii) Any new theory must be consistent with existing experiments (or astronomical observations); and, given it is going to be significantly different from either GR or QM, we expect to see highly non-trivial predictions of new phenomena, in many different areas of physics.

\vspace{2mm}

In the rest of this paper we discuss the CWL theory, formulated in an attempt to answer these demands. A useful way to begin, in explaining the reasoning leading to this theory, is to start by asking: what are the core features of QM and QFT that are essential, and hence cannot be dropped? And by the same token, what features of GR must we not sacrifice?

\subsubsection{Essential Features of QM}
 \label{sec:QM-essence}

In ref. \cite{stamp12} it was argued that the key feature of QM that should be kept is the idea of summation over paths. The value of this starting point is clear, for example, when we consider "interaction-free measurements" (as in, eg., 'negative result' experiments \cite{negR} or 'which path' experiments \cite{whichP} ). These show that the time evolution of a quantum system depends on what could have happened both along the paths it did follow, and also those it {\it did not} follow. A path integral formulation of QM makes this seems obvious, but it is less clear if we deal entirely in terms of $\langle \{ {\bf r}_j \}|\psi(t)\rangle$ (which is zero in regions where no paths are followed). In fact, interaction-free experiments simply exemplify the non-local character of QM, best seen in the Aharonov-Bohm effect - another interaction-free effect. This non-locality is most naturally understood in path integral language, where the fundamental role of the phase $\phi$ accumulated along a worldline becomes clear.

The advantages of path integrals are even more obvious in a relativistic theory.
In classical special relativity the worldline of a system is more fundamental than instantaneous events - the worldlines tie these events together to give meaning to the notion of a particle. In relativistic QM the paths in a path integral represent these worldlines; and standard relativistic QFT finds a natural non-perturbative formulation as a sum over dynamic field configuration paths. Note that the path integral formulation of QM and QFT is {\it not} in general equivalent to the wave-function formulation (a point first made forcefully by Morette-DeWitt \cite{morette}, and nicely exemplified in QM by the path integral formulation of fractional statistics, and in QFT by topological field theory).

Thus we will use the idea of QM paths as an essential starting point - wave-functions will no longer play a role. Note that {\it entanglement} between different systems, and {\it indistinguishability}, are also key parts of the experimental foundation of QM and QFT. In most texts in QM (but not QFT) these are described in terms of wave-functions. However, we can also describe them using a path integral formulation, and indeed this procedure has many advantages.

However, as discussed in ref. \cite{stamp12}, one element from standard QM that we {\it will} drop is the assumption of {\it independent paths}. Recall that in orthodox QM the standard Feynman result for the non-relativistic propagator $K_o({\bf r_2}, {\bf r_1}; t_2, t_1)$ of a single particle, between two spacetime 'end-points' $x_1 = ({\bf r_1}, t_1)$ and $x_2 = ({\bf r_2},t_2)$, takes the form
\begin{equation}
K_o(2,1) \;=\; \int^{2}_{1} {\cal
D} {\bf r}(\tau) e^{{i \over \hbar} S[2,1|{\bf r}(\tau)]}
 \label{Go-QM}
\end{equation}
with an {\it independent sum} over all paths ${\bf r}(\tau)$ extending between the end-points, and a weighting factor for each path involving an action $S[2,1|{\bf r}(\tau)] = \int^{t_2}_{t_1} d\tau L({\bf r}, {\bf \dot{r}})$, where $L$ is the classical Lagrangian. This sum embodies the QM superposition principle - and superposition is normally considered to be fundamental to QM. Nevertheless, starting in the next section, we will be setting up a consistent theory in which this principle is violated: we will introduce correlations between the paths, so that they no longer sum independently.

What this means is that we will simply generalize the propagator to \cite{stamp12,GR-Psi-L1,GR-Psi-1}:
\begin{equation}
  {\cal K}(2,1)\;=\; \sum_{n=1}^{\infty} \prod_{k=1}^n \int_1^2 {\cal D}{\bf r}_k \; \kappa_n[\{ {\bf r_k}(\tau) \}] e^{{i \over n \hbar} \sum_k S[2,1|{\bf r_k}(\tau)]  } \;\;\;\;\;
\label{DbbQ-1}
\end{equation}
where $S[2,1|{\bf r_k}(\tau)]$ is the action accumulated along the $k$-th path ${\bf r_k}(\tau)$ between the end-points (NB: we follow a practise here which is common in the literature, to write the exponent in the form $\sum_k F_k$, where $F_k$ depends on $k$, even though the sum in the exponent is redundant, given that we have a product over $k$ outside the exponential). The first term (with $n=1$) is just our standard result (\ref{Go-QM}) for conventional QM; the correlator $\kappa_n[{\bf r_1}, ... {\bf r_n}]$ in this sum then correlates $n$ different paths in the path integral, all with the same end-points. I have not specified a normalization for this path integral - this will be done below, when we do things properly.

Needless to say, these remarks do not yet make a theory - we have to do much more first. Let us now look at what is essential in GR.

\subsubsection{Essential Features of GR}
 \label{sec:GR-essence}

If one begins by assuming QFT to be fundamental, then it is natural to start from a spin-2 bosonic field, and show it must have an action of Einsteinian form \cite{deser70,wald86,feynman63}; the geometric view of spacetime then turns out to be secondary. However, one can also adopt a view in which both the metric field $g^{\mu \nu}(x)$ and the connection $\Gamma^{\lambda}_{\alpha \beta}(x)$ play a role quite different from that of all other fields, with a crucial geometric significance.

As we already saw, two key features of GR are (a) that Einsteinian GR is valid at low energies (but will break down at much higher energies); and (b) that once one quantizes matter, the quantization of the spacetime field cannot be avoided. But just what is it in low-energy GR that we must keep? We will demand two things:

\vspace{2mm}

(i) We must, at low energies, keep the fundamental notions of metric and affine connection, defined in terms of worldlines (or worldsheets for fields).  In general $g^{\mu \nu}(x)$ and $\Gamma^{\lambda}_{\alpha \beta}(x)$ are independent - only in conventional Einstein GR are they tied together by the Levi-Civita relation. Conventionally, one separates out those parts of $\Gamma^{\lambda}_{\alpha \beta}(x)$ dependent on $g^{\mu \nu}(x)$ by writing \cite{torsion}
\begin{equation}
\Gamma^{\lambda}_{\alpha \beta} \;=\; \bar{\Gamma}^{\lambda}_{\alpha \beta}(g) + K^{\lambda}_{\alpha \beta} + {1 \over 2}C^{\lambda}_{\alpha \beta}(g)
 \label{conn}
\end{equation}
where $\bar{\Gamma}^{\lambda}_{\alpha \beta}(g)$ is the Levi-Civita term, ie.,
\begin{equation}
\bar{\Gamma}_{\mu\alpha \beta} = \partial_{\alpha} g_{\mu \beta} + \partial_{\beta} g_{\mu \alpha} - \partial_{\mu} g_{\alpha \beta}
 \label{chris}
\end{equation}
and where the contortion $K^{\lambda}_{\alpha \beta}$ and the "con-metricity" $C^{\lambda}_{\alpha \beta}(g)$ are given in terms of the usual torsion tensor $S^{\lambda}_{\alpha \beta} = \Gamma^{\lambda}_{[\alpha \beta]}$ and non-metricity tensor $Q^{\lambda}_{\alpha \beta}(g) = \nabla^{\lambda}g_{\alpha \beta}(g)$ by
\begin{eqnarray}
K^{\lambda}_{\;\;\;\;\alpha \beta} &=& S^{\lambda}_{\;\;\alpha \beta} - S_{\alpha \beta}^{\;\;\;\;\lambda} - S_{\beta \alpha}^{\;\;\;\;\lambda} \nonumber \\
C^{\lambda}_{\;\;\;\;\alpha \beta} &=& Q^{\lambda}_{\;\;\alpha \beta} - Q_{\alpha \beta}^{\;\;\;\;\lambda} - Q_{\beta \alpha}^{\;\;\;\;\lambda}
 \label{KC-SQ}
\end{eqnarray}
and finally where the Riemann tensor $R^{\mu}_{\;\;\nu \alpha\beta}(\Gamma)$, also independent of $g^{\mu\nu}(x)$, is given by:
\begin{eqnarray}
R^{\mu}_{\;\;\nu \alpha\beta} \; = \;  \partial_{\alpha} \Gamma^{\mu}_{\;\; \nu \beta} &-& \partial_{\beta} \Gamma^{\mu}_{\;\; \nu \alpha} \nonumber \\ &+& \Gamma^{\mu}_{\;\; \lambda \alpha} \Gamma^{\lambda}_{\;\; \nu \beta} - \Gamma^{\mu}_{\;\; \lambda \beta} \Gamma^{\lambda}_{\;\; \nu \alpha} \;\;\;\;
 \label{Riemann}
\end{eqnarray}
To get back Einstein GR one puts both $S^{\lambda}_{\alpha \beta}$ and $Q^{\lambda}_{\alpha \beta}$ to zero, and the Levi-Civita relation (\ref{chris}) then fixes $\Gamma$ in terms of $g$.

\vspace{2mm}

(ii) We will assume that the "Einstein Equivalence principle" (EEP) is valid - we will in fact extend this principle in a rather specific way to the CWL theory. Thus worldlines for "test particles" (and worldsheets for test fields) are assumed independent of the nature of the underlying particle or field; and freely falling particles or freely accelerating fields can be transformed locally (apart from tidal forces) to a local inertial frame. Put simply, all fields and particles couple to gravity in the same way, and inertial and gravitational masses are the same.

\vspace{2mm}

This then implies that the non-metricity $Q^{\lambda}_{\;\;\alpha \beta} = 0$. The EEP as defined above does not exclude torsional degrees of freedom, but it does exclude any terms, explicitly dependent on the connection, from appearing in the action. One can have a more restricted EEP, in which spin degrees of freedom are allowed to see the gravitational field differently from scalar or EM fields - such theories emerge if we keep $g^{\mu \nu}(x)$ and $\Gamma^{\lambda}_{\alpha \beta}(x)$ independent (the Einstein-Cartan-Kibble-Sciama theory \cite{ECKS} being the best-known).

In what follows we will assume 4-dimensional gravity; and we will ignore higher-derivative theories. However, we will make a key change, outside the usual framework of classical GR. Once we go over to a quantum theory, in which worldlines becomes paths in a path integral, we will need to generalize the EEP to encompass correlated worldlines. We will then see how this allows us to derive the form of these correlations in the CWL theory.


\section{Derivation of the Correlated Worldline Theory}
 \label{sec:CWT-arg}


As it stands, the idea of correlating paths is no more than an interesting observation. It yields a formal framework with the mathematical advantage of being extremely "open" - we can essentially choose any form for the correlators $\kappa_n[{\bf r_1}, ... {\bf r_n}]$ that we like. However it is physically empty until we can specify, using physical arguments, what these correlators may be.

The key steps in formulating CWL theory, already sketched in ref. \cite{stamp12}, then consist in (i) setting up the mathematical framework for correlated worldlines, and then (ii) fixing the form of these correlations using arguments coming from gravitation.

In what follows we discuss the physical arguments first, and then set up the theory in terms of a specific set of postulates.

\subsection{Rationale for the CWL Theory}
 \label{sec:CWL-rat}

Why would one want to have correlations between different Feynman paths? In what follows a set of physical arguments is adduced to justify this idea. We begin by asking how one should think about Feynman paths in GR (a topic which has of course been discussed repeatedly in previous work \cite{mueller13}), and then, in a departure from previous analyses, we discuss why they ought to be correlated.

\subsubsection{Reintroducing the metric and the affine connection}
 \label{sec:CWL-metric-aff}

We will assume, to simplify things, that the basic objects in the theory are particle worldlines - even though it is more correct to start from fields. The discussion for quantum fields is technically more demanding \cite{GR-Psi-1}.


\begin{figure}
\vspace{0.2cm}
\includegraphics[width=\columnwidth]{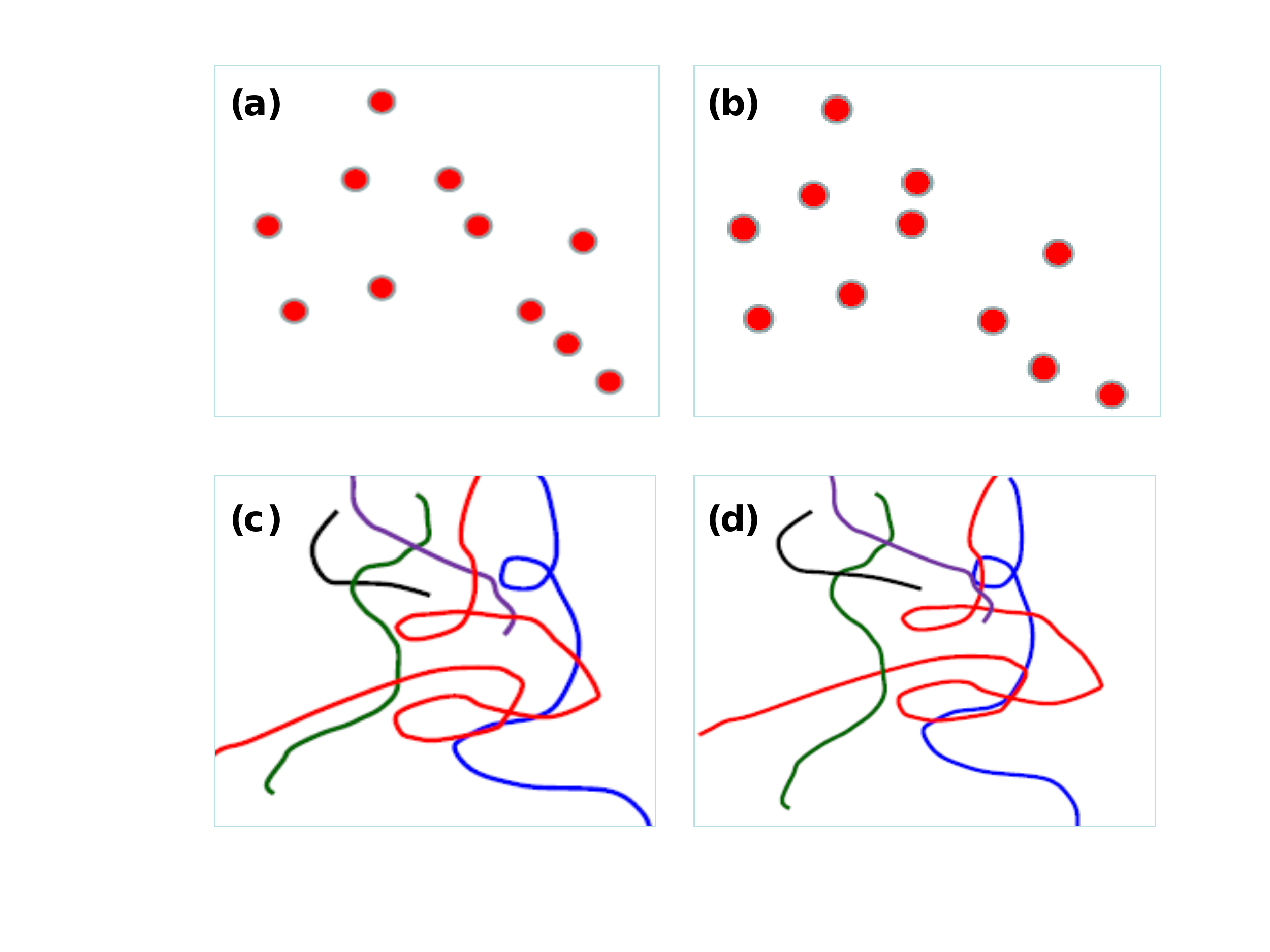}
 \vspace{-0.9cm}
\caption{Comparison of 2 different configurations in spacetime. In (a) and (b) we see 2 different configurations, each of which has 11 different "points" or "spacetime coincidences". Although they look fairly similar, there is no unique way of matching them together. In (c) and (d) we compare 2 different sets of 5 worldlines. If they are labeled by different colours, we can match them, up to diffeomorphism transformations; if they are not labeled (ie., all of the same colour) we have no way of matching them.  }
 \label{fig:NJP-F3}
\end{figure}


Consider then the situation shown in Fig. \ref{fig:NJP-F3}. We imagine 2 sets of quantum worldlines, or Feynman paths, labeled by $W_1$ and $W_2$, which here could represent a set of $N$ particles in 2 different configurations (the generalization to a set of matter fields in 2 different configurations should be fairly clear). In conventional GR each set would exist on separate manifolds ${\cal M}_1$ and ${\cal M}_2$. If we wish to compare different worldlines on a single manifold, we still have all the ambiguities involved in diffeomorphism invariance, which are dealt with in GR using connection and metric fields on this manifold. However a superposition of the two different configurations, existing separately on ${\cal M}_1$ and ${\cal M}_2$, would, as we already saw, be quite meaningless. \cite{penrose96,penroseRR}.

For those unimpressed by these mathematical ambiguities, consider the following simple question: how are we supposed to measure how similar $W_1$ is to $W_2$, once we have {\it removed both the "background spacetime" and all labeling of the different particles}? The latter step, ie., the removal of the labels, is not part of orthodox GR - however it will be necessary if we are to include quantum-mechanical indistinguishability in our picture. It is obvious, if we now compare the 2 pictures, that there is no unique way to do this (one could, eg., try a 'least squares fit', if all the lines seem to be close to each other, but this is clearly completely arbitrary). We find ourselves again with the problem of how to superpose spacetimes - but now we are discussing it in terms of quantum worldlines, ie., Feynman paths.

We therefore start from the view that {\it quantum phase is fundamental}, and ask how, in a theory of quantum gravity, are we going to (i) define the quantum phase for one of the physical systems moving along one specific path; and (ii) define the relationship between the quantum phase on one worldline, and that on the other?

\vspace{2mm}

{\bf (i) A single worldline:} Suppose we are travelling along a "test worldline", with no awareness of any other worldines, either for the same particle or for any other particle. How should we parametrize the accumulated phase?

Clearly we need an object which is capable of sensing, at the very least, the rotational and perhaps torsional properties of spacetime. We begin by introducing an "internal clock", to define our position along the worldline. Quantum mechanics provides with such a clock in terms of the internal phase associated with an elementary particle (eg., an electron) of mass $m$; this phase is $mc^2\tau/\hbar$, defining for us a proper time $\tau$.

We then introduce an "internal metric" which allows us to refer changes in phase at some point along the worldline to the value at an initial point; and we introduce a connection variable $\Gamma^{\alpha}_{\mu\nu}$ to define the gradients of this phase. The most common way to do this is by introducing a set of tetrads $\{ e^{\mu}_j \}$ on the worldline, defining a local orthonormal set of coordinates which move along the worldline, in terms of some other fixed set of spacetime coordinates with basis vectors ${\bf e}^{\mu}$. We do this here for ease of exposition, even though it partially disguises what is going on.

\begin{figure}
\vspace{0.2cm}
\includegraphics[width=\columnwidth]{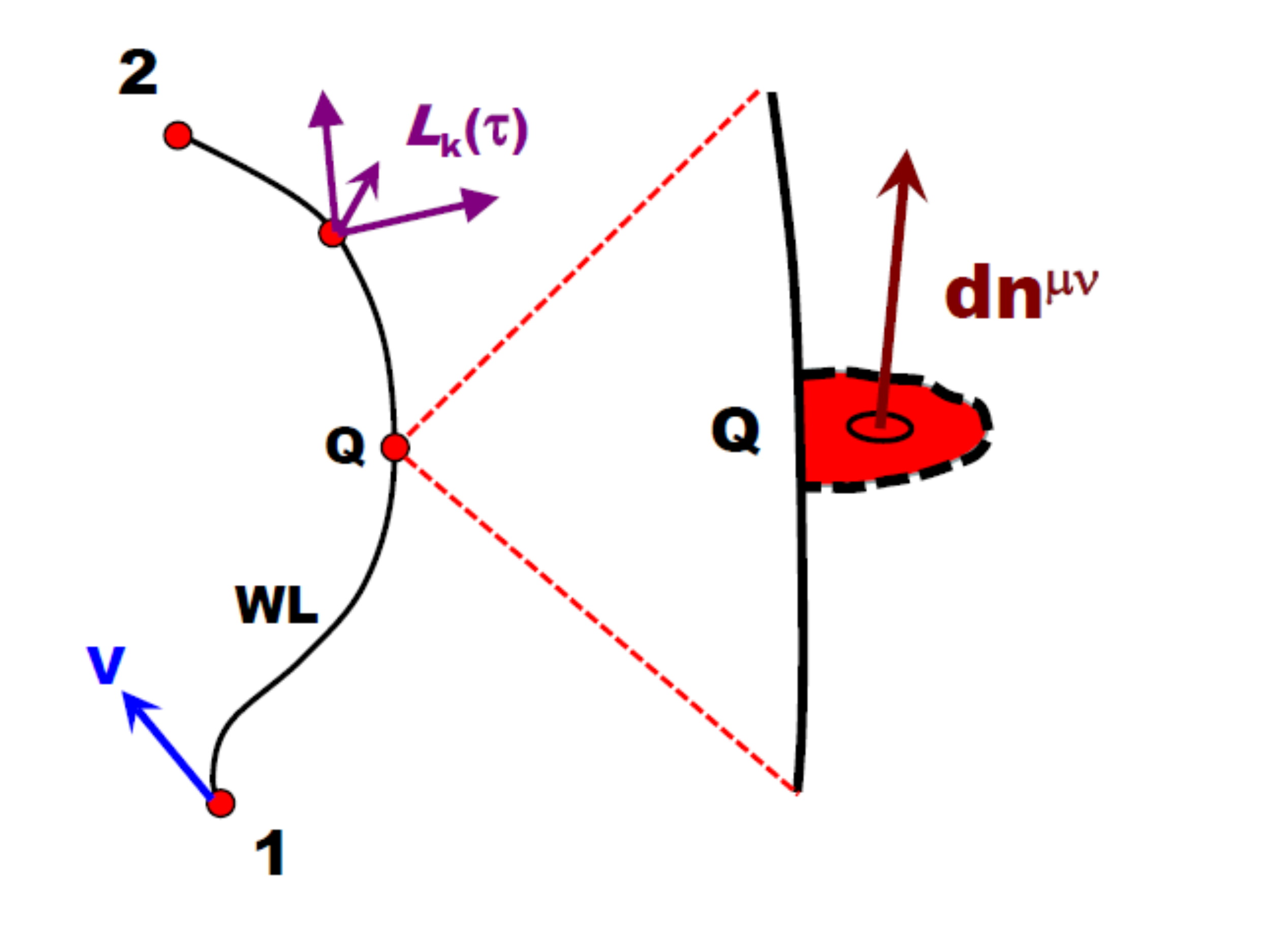}
 \vspace{-0.9cm}
\caption{At left, a worldline $q(\tau)$ for a particle, defined between the end-points 1 and 2. The position along the worldline is defined by a proper time $\tau$, and $L_k(\tau)$ is an internal coordinate frame moving with the particle. It can be compared with the tangent vector $V(1)$ defined at the endpoint 1 of the path. At right, we blow up a small section of the worldline, and imagine an infinitesimal detour in the path at a point $Q$, with an oriented area element $dn^{\mu\nu}$ enclosed. }
 \label{fig:NJP-F4}
\end{figure}

Consider then a worldline path $q(\tau)$ along which our test object is moving (compare Fig \ref{fig:NJP-F4}); we wish to know how the phase accumulates along the worldline, and how it may change of we change the worldline.

\begin{widetext}

In standard GR, strongly supported by experiment, we can define an operator acting along the path $q(\tau)$ between 1 and 2, given by
\begin{equation}
\hat{U}_{21}[q] \;=\; \hat{T} \exp {i \over \hbar} \int  dx^i  \int^{\tau_2}_{\tau_1} d\tau' \delta(x^i - q^i(\tau')) \; [e^{\mu}_k \partial_{\mu} - \Gamma_{\mu\nu}^{\alpha}(x) e^{\mu}_i e^{\nu}_j {\partial \over \partial e^{\alpha}_j} ]
 \label{U-path}
\end{equation}
and this operator yields this phase (in the form of the argument of the operator). Here $\hat{T}$ orders the infinitesimal parts of the exponential, as we go along the path; and no metric structure has yet been attributed to the spacetime, to which we have given the coordinate system $x^j$.

\end{widetext}

We can also make an infinitesimal distortion of the path at some point, at a point $Q(s)$ along $q(\tau)$, which encloses a small extra area $dn^{\mu\nu}(Q)$ (whose circumference we define using the detour in the path). Functionally differentiating (\ref{U-path}), we then find that
\begin{equation}
\delta_Q \hat{U}_{21}[q] \; =\;  - R_{\mu \nu}(Q) \;\hat{U}_{21}[q] \; dn^{\mu \nu}(Q)
 \label{dU-path}
\end{equation}
where the curvature tensor $R_{\mu \nu} = R^{\alpha}_{\mu \nu \alpha}$, with $R^{\alpha}_{\mu\nu\beta}$ defined in terms of the connection by (\ref{Riemann}).


The general idea of this thought experiment is that an observer looking at this phase detector would discover that the best way to parametrize the way in which the accumulated phase varied, as one changed at will the path that it followed, would be in terms of an affine connection. This is at least one consistent way of defining the quantum properties of the Feynman path.


\begin{figure}
\vspace{-0.1cm}
\includegraphics[width=\columnwidth]{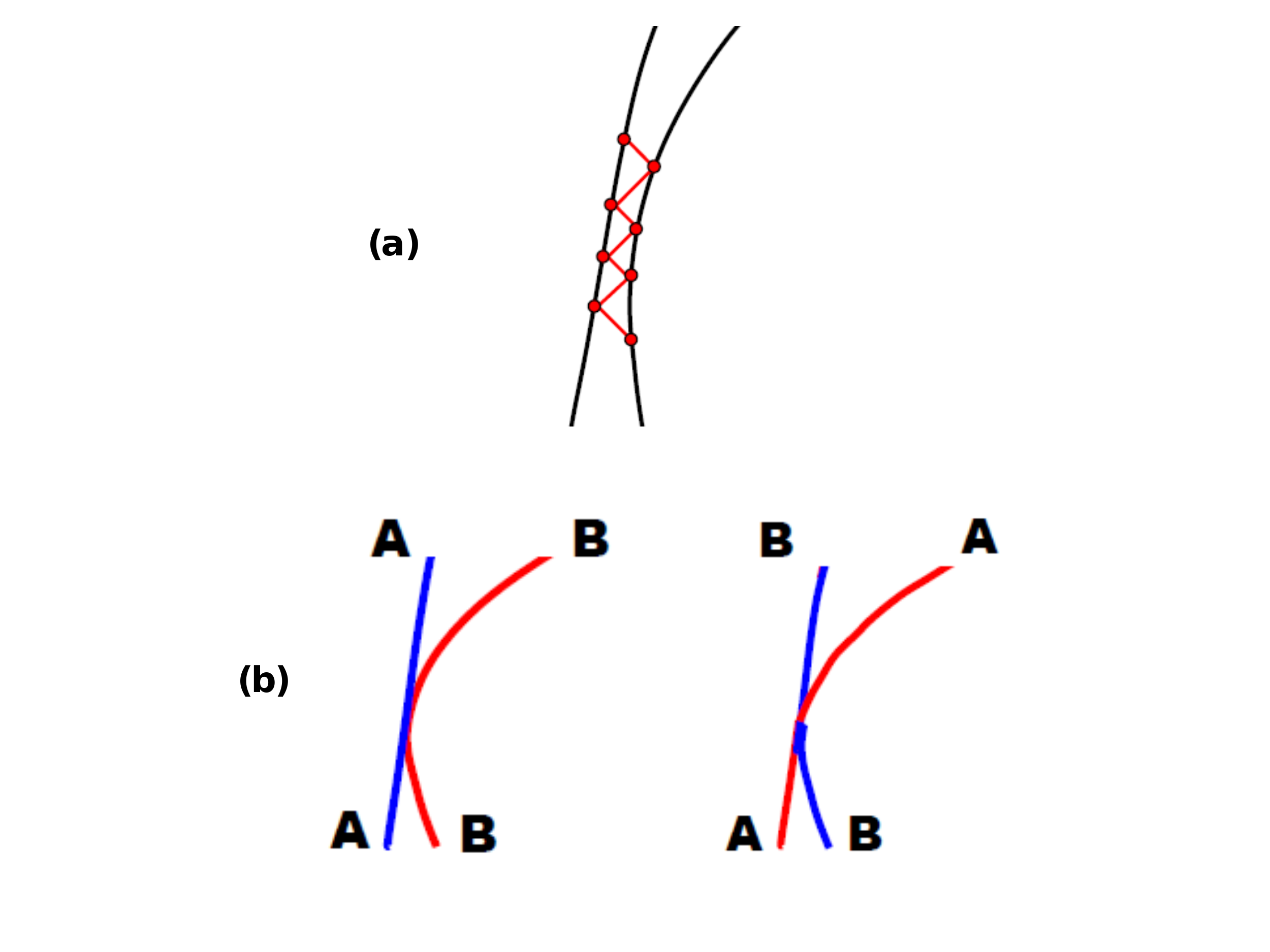}
 \vspace{-0.8cm}
\caption{Communication between 2 paths. (a) different worldlines $A$ and $B$ communicate via exchange of light signals, allowing a definition of $g^{\mu\nu}$. (b) 2 worldlines for indistinguishable particles $A$ and $B$ - there is no way of distinguishing between the 2 processes shown. }
 \label{fig:NJP-F5}
\end{figure}


\vspace{2mm}

{\bf (ii) worldlines for 2 different particles:} Quantum phase is usually defined by either comparing two paths or closing a single path (we have avoided this necessity by using an internal clock). In QM one also has to specify whether one is dealing with 2 paths for the same particle, or 2 paths for 2 different particles.

Consider first 2 different particles (Fig. \ref{fig:NJP-F5}). We communicate between the 2 worldlines using light signals, allowing a determination of the relation between the phases on the 2 lines chronometrically. We can then introduce the metric, in a way standard in GR \cite{synge60,LL-fields}, by timing light signals between the paths; we imagine 2 particles approaching each other very closely ("close" being defined by the proper time measured for signals to go back and forth). One can then extract all elements of the metric tensor $g^{\mu\nu}(x)$, within a constant factor, in the neighbourhood of the point $x$ where the particles would "touch" ($x$ is thus a "spacetime coincidence"). The metric tensor is now defined in terms of the measurements of intervals between 2 separate points, one on each worldline, defined in terms of relative phases between them.

\subsubsection{A Generalized Equivalence Principle}
 \label{sec:CWL-EEP}

In the above discussion, no relationship has yet appeared between the connections used for each different worldline (which relate to their individual phases), and the metric defining the distance between the two worldlines. We now introduce the principle of equivalence, to see how this changes things.

\vspace{2mm}

{\bf (i) Two particles}: The principle of equivalence makes no distinction between (a) two worldlines for 2 different and distinguishable particles, and (b) two worldlines for two indistinguishable particles. As far as gravitation is concerned, these two cases look exactly the same. Now in case (a) we describe the relationship between the 2 worldlines using the metric, in the usual way. However we notice that in case (b) we can imagine an exchange process in which the 2 particles swap roles, and so the 2 worldlines actually refer to the same particle. In this latter case the relationship between the 2 worldlines can then described using the connection, using eqtn. (\ref{dU-path}). Without going through a formal demonstration, we have established a relationship between metric and connection, which is actually just the Levi-Civita relation in (\ref{chris}).

An immediate corollary of this is that the self-interaction, via gravitation, of a particle has exactly the same form as the gravitational interaction between 2 particles of the same mass.

\vspace{2mm}

{\bf (ii) Two worldlines for a single particle:} Now we can complete the argument. We have just seen how the principle of equivalence must apply to two different but indistinguishable particles. Consider now 2 paths for the {\it same} particle. In a path formulation of the dynamics, there is no obvious reason why the gravitational field should relate 2 wordlines referring to 2 indistinguishable particles any differently from the way it relates 2 worldlines referring to the {\it same particle}. If we take this remark seriously, then we must now extend the equivalence principle to cover two different Feynman paths for the same particle. (clearly this argument needs refinement to deal with fermionic fields; this is left for another paper \cite{GR-Psi-1}).

We therefore arrive at a key result - the gravitational correlation established between the worldlines of two different particles will be precisely the same as that established between 2 paths for the same particle. This immediately signifies a breakdown of the superposition principle in QM, of precisely the kind we were looking for. Let us now see how to build a theory from this observation.

\subsection{Formal Postulates of the CWL Theory}
 \label{sec:CWL-post}

To reformulate these physical arguments more clearly, I outline 3 postulates, which will then allow us to write down a formal CWL theory. The first 2 postulates determine the general mathematical structure of the theory; the third postulate, of a physical nature, then fixes it uniquely.

\subsubsection{The first two Postulates: Mathematical Structure}
 \label{sec:post1+2}

Our first two postulates determine the structure of the theory:

\vspace{2mm}

{\it Assumption 1}: We assume that the worldlines for particles (or worldsheets for fields) are fundamental, along with the phase $\phi$ accumulated along a worldline; we assume $\phi = S/\hbar$, where $S$ is the action accumulated along the worldline, as determined by the classical Lagrangian.

As a corollary to this we will assume that spacetime itself will in the first instance be defined in terms of the worldlines or worldsheets of appropriate objects.

\vspace{2mm}

{\it Assumption 2}: In contrast to orthodox QM, we allow correlations between the worldlines; in QM, one sums over these independently, but here we allow a violation of the superposition principle.

Notice also that nowhere have we mentioned operators, or Hilbert space, or measurements. A detailed discussion of quantum measurements, as "just another physical process" (albeit one of a specific kind) is reserved for another paper \cite{GR-Psi-1}; but a sketch of how these are incorporated into the CWL theory is given in section \ref{sec:CWL-QM-mmt} below.

What Assumptions 1 and 2 deliver us is the idea that separate paths are allowed to be correlated, by a set of correlators $\kappa_r[q_1, q_2, ...q_r]$ between the worldlines $\{ q_k \}$. Now, without saying anything yet about the physical mechanism causing the correlations, let us first summarize the structure of the theory that then follows.

In introductory courses on conventional QFT \cite{ryder+al} one learns that the entire theory is determined, starting from the generating functional ${\cal Z}[J]$, by functional differentiation of ${\cal Z}[J]$ with respect to $J$. This gives all the correlators of the theory, in terms of which all experimental properties can be determined. It turns out that the easiest way to write down the formal structure of CWL theory is by a simple generalization of this procedure.

We begin by considering a single relativistic particle, and define a generating functional ${\cal Q}[j] \;=\; {\cal Z}_o[j] + \Delta {\cal Q}[j]$, in the presence of an external current $j(\tau)$ acting on the particle, as:
\begin{eqnarray}
{\cal Q}[j] &=& {1 \over \cal N}\; \sum_{n=1}^{\infty} \prod_{k=1}^n \int {\cal D}q_k \; \kappa_n[\{q_k \}] e^{{i \over n \hbar}\sum_k (S[q_k] + \int j q_k)   } \;\;\;\;\; \nonumber \\
 &=& \sum_{n=1}^{\infty} Q_n[j]
\label{DbbQ}
\end{eqnarray}
where $S[q_k]$ is the action accumulated along the $k$-th path $q_k(\tau)$. The normalization factor ${\cal N}$ is just
\begin{equation}
{\cal N} = \sum_{n=1}^{\infty} \prod_{k=1}^n \int {\cal D}q_k \; \kappa_n[\{q_k \}] e^{{i \over n \hbar} \sum_k S[q_k]  }
 \label{N1}
\end{equation}
and we are again using the convention where we write the sum $\sum_k S[q_k]$ in the exponent, even though the summation is superfluous given the product $\prod_k$ outside the exponent.

The functional ${\cal Q}[j]$ is precisely what we need as the generalization of the standard partition function ${\cal Z}[j]$. If the first term (with $n=1$) from (\ref{DbbQ}) is taken for the moment to have $\kappa_1[q] = 1$, we then have
\begin{equation}
Q_1[j] \; \rightarrow \; {\cal N}_o^{-1} \int {\cal D}q \; e^{{i \over \hbar} (S[q] + \int j q)   } \; \equiv \; {\cal Z}_o[j]
 \label{Z-o}
\end{equation}
which we see is just the usual generating functional ${\cal Z}_o[j]$ for a free relativistic particle, in conventional QM; and ${\cal N}_o = \int {\cal D}q \; e^{{i \over \hbar} S[q]   }$.

\begin{figure}
\vspace{0.4cm}
\includegraphics[width=\columnwidth]{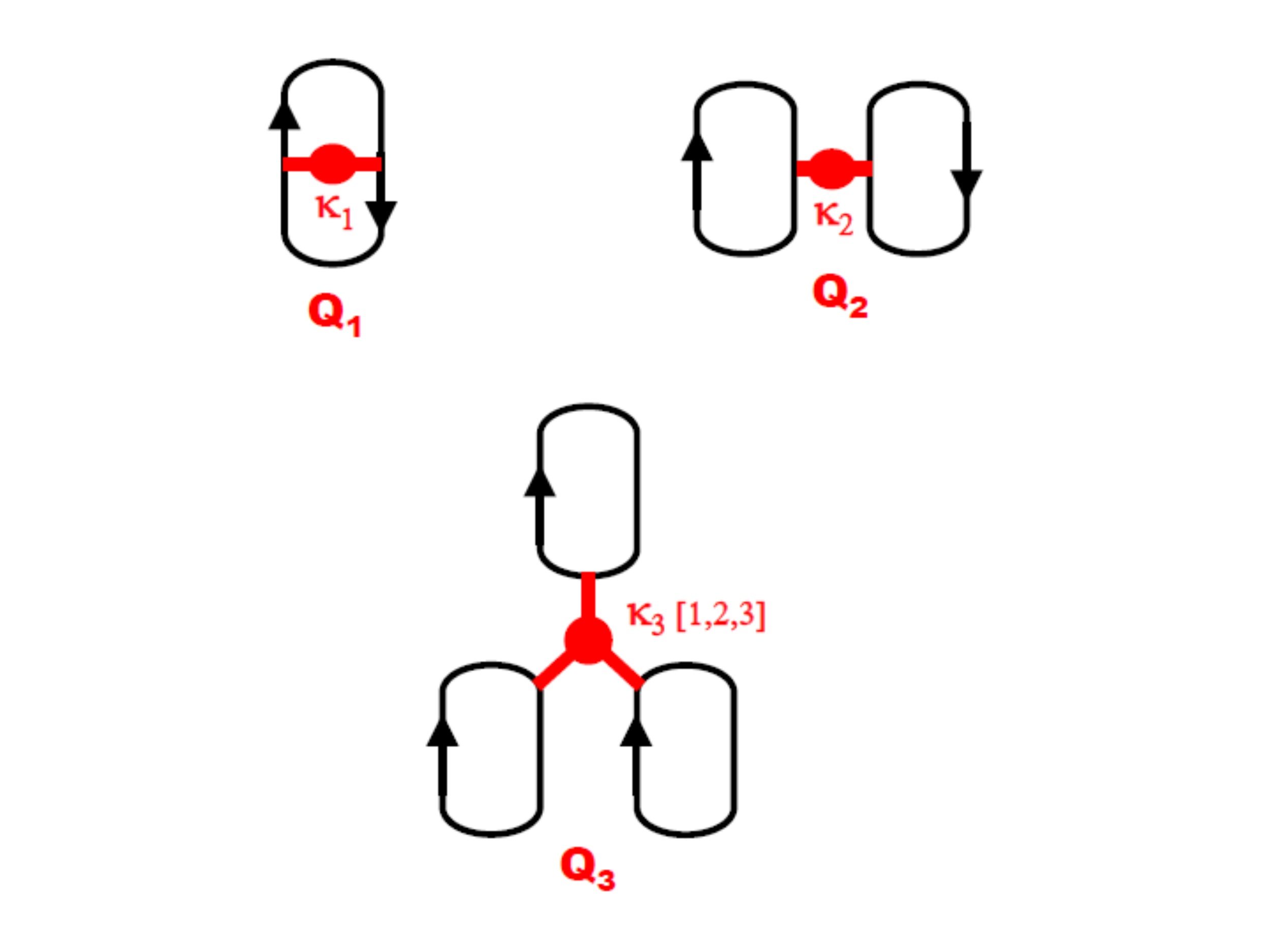}
 \vspace{-0.6cm}
\caption{Contributions to the ring sum in eqtn. (\ref{DbbQ}). We show the first 3 contributions $Q_1[j], Q_2[j]$, and $Q_3[j]$. In eqtn. (\ref{Z-o}), which describes conventional QM without gravitation, the correlator $\kappa_1 = 1$, and $\kappa_n = 0$ for all $n>1$. }
 \label{fig:NJP-F6}
\end{figure}

The correlator $\kappa_n[q_1, ...q_n]$ in ${\cal Q}[j]$ then correlates $n$ different paths in the path integral. It will be convenient in what follows to choose a 'ring path' extending around a loop from $t=-\infty$ to $t=+\infty$ (ie., we join at $t=-\infty$ the 2 legs of the usual Keldysh/Kadanoff-Baym/Schwinger or 'in-in' path \cite{keldysh}); then ${\cal Q}[j]$ can be represented as a 'ring sum', shown in Fig. \ref{fig:NJP-F6}. Recall that in the standard "in-in" formalism one has a path extending from $t=-\infty$ to $t=+\infty$ and then back to $t=-\infty$; here we simply close the curve (as was done by Kadanoff and Baym) at $t=-\infty$, by displacing the in and out lines by an imaginary time $-i\beta$, where $\beta = 1/kT$. One can if one likes think of this as a path on a cylinder, with circumference $2\pi\beta$, with the path making a complete circuit of the cylinder in order to complete the ring \cite{BlochdeD}.


At the moment the introduction of these rings should be considered as merely a formal device, used to make calculations tractable. However we note that it means that we always deal with {\it even} numbers of lines, in which forward and backward propagators are paired.

Ultimately it makes no sense to consider relativistic particles when dealing with gravity - one has to introduce quantum fields. The multi-ring construction is easily generalized to quantum fields. Consider, eg., a scalar field $\Phi(x)$ coupled to a source $J(x)$; one then has a 'correlated worldsheet' picture, in which the ring sum becomes
\begin{eqnarray}
{\cal Q}[J] &=& {1 \over \cal N}\;\sum_{n=1}^{\infty} \prod_{k=1}^n \int {\cal D}\Phi_k \; \kappa_n[\{\Phi_k \}] \nonumber \\ && \;\;\;\;\;\;\;\;\;\;\;\;\;\;\;\;\;\;\;\;\;\;\; \times \; e^{{i \over n\hbar} (S[\Phi_k] + \int d^4x J(x) \Phi_k(x))   } \;\;\;\;\;\;\;
\label{DbbQ2}
\end{eqnarray}

Let us emphasize again that these two postulates are purely formal - all they do is set up a mathematical framework.

\subsubsection{Postulate 3: the Role of Gravity}
 \label{sec:post-3}

We have already given the arguments surrounding our physical idea of what the correlators $\kappa_n$ should be; they are summarized by the following postulate:

\vspace{2mm}

{\it Assumption 3}: The correlation between the different branches of a superposition is established by the gravitational field itself, acting via the set of correlators $\kappa_r[q_1, q_2, ...q_r]$ between worldlines; and the principle of equivalence applies equally to all the correlators.

\vspace{2mm}

This last assumption on the CWL theory is far-reaching - indeed it fixes the form of the theory uniquely. We can think of it as a "generalized EEP", or generalized equivalence principle.

To see how this works, let's go back to our free relativistic particle, and consider what happens when we couple it to gravity. The action for the particle is then well-known:
\begin{eqnarray}
S[q, g^{\mu \nu}] &=& -\int ds {m \over 2} g_{\mu \nu}(q(s)) \; \dot{q}^{\mu} \dot{q}^{\nu} \nonumber \\
&=& -\int d^4x \int ds {m \over 2} g_{\mu \nu}(x) \dot{x}^{\mu} \dot{x}^{\nu} \delta(x - q(s)) \;\;\;\;\;\;\;
 \label{freeP}
\end{eqnarray}
where the derivatives are with respect to the proper time $s$. In the absence of the correlations between rings we have just a single ring - this is just conventional QM, for a particle coupled to the background spacetime metric. Then we just recover the conventional generating functional ${\cal Z}[j]$ for the dynamics of this particle, ie., ${\cal Q}[j] \rightarrow  {\cal Z}[j]$, where
\begin{eqnarray}
{\cal Z}[j] &=& {1 \over {\cal N}} \int {\cal D}g^{\mu\nu} e^{{i \over \hbar} S_G[g^{\mu \nu}]} \Delta[g^{\mu \nu}] \nonumber \\
&& \;\;\;\;\;\;\;\;\;\;\;\;\;\;\;\;\; \times  \int {\cal D}q \; e^{{i \over \hbar} [S[q] + \int ds j(s) q(s)]   } \;\;\;\;\;\;
 \label{Z-part}
\end{eqnarray}
in which we now have a functional integral over all metrics consistent with the restriction to energies $\ll E_p$, and where we have introduced a Faddeev-Popov \cite{faddeevP} determinant $\Delta[g^{\mu \nu}]$, which serves to properly normalize the path integration (by factoring out all diffeomorphism-equivalent metric configurations). The vacuum-vacuum amplitude is ${\cal N}$, and  $S_G[\tilde{\mathfrak{g}}^{\mu \nu}(x)]$ is the usual Einstein action, viz.:
\begin{equation}
S_G \;=\; {1 \over \lambda^2} \int d^4x \; [\tilde{\mathfrak{g}}^{\mu \nu} R_{\mu \nu} - {1 \over 2 \alpha} (\partial_{\mu} \tilde{\mathfrak{g}}^{\mu \nu})^2]
 \label{S-G}
\end{equation}
where we have included a gauge-fixing term consistent with the Faddeev-Popov determinant.

Eqn. (\ref{Z-part}) simply tells us that in standard quantum gravity, we have the result
\begin{eqnarray}
\kappa_1[q] &=& \int {\cal D}g^{\mu\nu} e^{{i \over \hbar} S_G[g^{\mu \nu}]} \Delta[g^{\mu \nu}] \nonumber \\
\kappa_n[q] &=& 0 \;\;\;\; (\forall \; n > 1) \;\;\;\;\;\; (standard \; Q. \; Gravity) \;\;\;\;\;\;\;
 \label{QGrav}
\end{eqnarray}

However if we now introduce Assumption 3, it is immediately clear from our generalized principle of equivalence that the correlation between different ring paths must be the same as that inside a given ring - thus we must generalize (\ref{QGrav}) to:
\begin{eqnarray}
\kappa_n[q] &=& {1 \over n!} \int^{\bf \prime \prime} {\cal D}g^{\mu\nu} e^{{i \over \hbar} S_G[g^{\mu \nu}]} \Delta[g^{\mu \nu}] \;\;\;\;\;\;\;(\forall \; n > 0) \;\; \nonumber \\
&& \;\;\;\;\;\;\;\;\;\;(CWL \;\; theory)
 \label{kappa-n}
\end{eqnarray}

This is our key result - it fixes completely the form of the theory, and will allow us to calculate any physical quantity from it. The insertion of the factor $1/n!$ is a matter of choice (one can just as easily absorb it into the Faddeev-Popov determinant). From a physical point of view it emphasizes the indistinguishability of the different paths in the sum over $n$. The double prime symbol above the integral indicates formally that in summing over multiple paths ($n$ different paths for $\kappa_n$) we exclude any identical paths \cite{repeatP}.

Notice the extent to which the generalized principle of equivalence determines things - we have no choice, once we have accepted it, in accepting the form in (\ref{kappa-n}) for $\kappa_n[q_q,q_2,...q_n]$.

From (\ref{kappa-n}) we can now write down the complete form for the generating functional. Thus, eg., for our relativistic particle we now have
\begin{eqnarray}
{\cal Q}[j] &=& \sum_{n=1}^{\infty} \prod_{k=1}^n  \int^{\bf \prime \prime} {\cal D}\tilde{\mathfrak{g}}^{\mu \nu} \int {\cal D}q_k  \nonumber \\
&\times& {1 \over n!} e^{i S_G/\hbar} \; \Delta[\tilde{\mathfrak{g}}^{\mu \nu}] \;  e^{{i \over n\hbar} \sum_k (S[g, q_k] + \int ds \;j(s) q_k(s))  } \;\;\;\;\;\;
 \label{Qpart-gen}
\end{eqnarray}
where we see that the correlator $\kappa_n$ acts on each of the paths $q_k(s)$ via the presence of the metric $g^{\mu\nu}(x)$ in the particle action (\ref{freeP}), and then through the functional integration over all configurations of $g^{\mu\nu}(x)$ consistent with  our assumption of a low-energy theory. We discuss below how this generalizes to a quantized field.

Notice that $\kappa_n[q_q,q_2,...q_n]$ is not explicitly dependent on the particle path - this is because we have assumed a form for the principle of equivalence in which the matter action depends only on the metric, and not on the connection. However this does {\it not} mean that $\kappa_n$ has no effect on the dynamics - this is because the particle action itself depends on the metric, and from (\ref{freeP}) we have
\begin{equation}
S[q_k, g^{\mu \nu}] \;=\; -\int ds {m \over 2} g_{\mu \nu}(q_k(s)) \; \dot{q}_k^{\mu} \dot{q}_k^{\nu}
 \label{S-p-k}
\end{equation}
with the metric $g_{\mu \nu}(x)$ evaluated on the path $q_k(s)$.

\begin{widetext}

We can gain more insight on this by asking what form the correlator would have to assume if we instead used the more restricted principle of equivalence, in which the connection $\Gamma^{\alpha}_{\mu\nu}$ and the metric density $\tilde{\mathfrak{g}}^{\mu \nu}(x)$ are allowed to be independent fields. In this case we need to define a different connection function for each particle worldline; and so we now must have:
\begin{equation}
\kappa_n [\{ q_k \}] \;\;=\;\; {1 \over n!} \int^{\bf \prime \prime} {\cal D}\tilde{\mathfrak{g}}^{\mu \nu}(x) \int^{\bf \prime \prime} {\cal D} ^{(k)}\Gamma^{\lambda}_{\alpha \beta}(x) \;
e^{{i \over \hbar} S^{(k)}_G} \Delta[\tilde{\mathfrak{g}}^{\mu \nu}(x),\; ^{(k)}\Gamma^{\lambda}_{\alpha \beta}(x)]
  \label{kappan2}
\end{equation}
where $^{(k)}\Gamma^{\lambda}_{\alpha \beta}(x)$ is the connection for the $k$-th path or field configuration of the system under consideration, where $\Delta[\tilde{\mathfrak{g}}^{\mu \nu}(x),\; ^{(k)}\Gamma^{\lambda}_{\alpha \beta}(x)]$ is another Faddeev-Popov determinant, now generalized to deal with independent connection and metric density configurations, and where the action $S^{(k)}_G$ is
\begin{equation}
S^{(k)}_G = \int d^4x \tilde{\mathfrak{g}}^{\mu \nu}(x) R^{(k)}_{\mu \nu}(x)
 \label{S-kG}
\end{equation}
plus any gauge-fixing terms. The curvature tensor $R^{(k)}_{\mu \nu}(x)$ for the $k$-th path is given in terms of $^{(k)}\Gamma^{\lambda}_{\alpha \beta}(x)$ by
\begin{equation}
R^{(k)}_{\mu \nu}(x) \;\; = \;\; (\partial_{\mu}\; ^{(k)}\Gamma^{\alpha}_{\alpha \nu} - \partial_{\nu} \; ^{(k)}\Gamma^{\alpha}_{\alpha \mu}) \;\; + \;\; ^{(k)}\Gamma^{\alpha}_{\beta \mu}  \; ^{(k)}\Gamma^{\beta}_{\alpha \nu} - \; ^{(k)}\Gamma^{\alpha}_{\nu \mu} \; ^{(k)}\Gamma^{\beta}_{\alpha \beta}
 \label{R-k-uv}
\end{equation}
which is, as we expect, independent of the metric density $\tilde{\mathfrak{g}}^{\mu \nu}(x)$. Notice now the the connection and the curvature for a given particle are dependent on which worldline we take - but the metric density, which relates worldlines, is considered to be independent of these.
\end{widetext}

Now in such a "metric-affine" CWL theory, as we have seen already, we would have to interpret the connection variables $^{(k)}\Gamma^{\alpha}_{\mu \nu}$ as encapsulating all information about the internal phase of the particle, and how it depends on spacetime - whereas the metric density $\tilde{\mathfrak{g}}^{\mu \nu}(x)$ would give information on the relative phases of the 2 worldlines.

However as we have also seen, it makes no sense, given that we are comparing worldlines of the same particle, to apply such a restricted form of the principle of equivalence. We then see that the role of the correlator in its original form (\ref{kappa-n}) is precisely to define the relative phase of two paths in a superposition. To put it another way: information comparing what one path is doing with that of another, in a quantum theory, is communicated between the paths by the metric density function (as well as between different parts of the same path).


\section{Correlated Worldline Theory: Basic Structure}
 \label{sec:CWL-structure}


We have now fixed the form of the theory - but we also need to see how it works. Indeed, when looking at any theory, one first wants to get an idea of what 'shape' it has - how one relates the formal structure to the physical properties, how things change with time, what are the important limiting cases, and how to think intuitively about the theory and its consequences.

Now QM and GR have very different 'shapes'. In conventional QM we start from a very peculiar formal structure - state vectors in infinite-dimensional Hilbert space, with an operator structure which is supposed to relate this to "measurements" and external observers, and with a formally arbitrary (and physically nonsensical) divide between "system" and "apparatus". However the theory is a linear one, which makes it easy to understand the time evolution of the state vectors in terms of a simple differential equation, and to think intuitively in terms of 'state superpositions', interference, and so on. The non-linear character of QFT requires a more complex apparatus, because of interactions, but so long as these are understood using perturbative expansions, the basic form or 'shape' of the theory doesn't change too drastically.

GR is radically different. The theory is fundamentally non-linear, and this changes everything - all the intuition acquired in dealing with a linear theory must be thrown away. A few examples suffice to make this point - the non-localizability of energy, the peculiar way gravitational waves propagate over long times, and the complete nonsense one gets if one calculates the vacuum energy of a quantum field in a curved spacetime by adding the zero-point energies of each mode \cite{wald04}. In GR one must think globally - the solutions to the field equations cannot be understood any other way - and this is what makes it hard to develop an intuitive feeling for the theory, even though its classical ontological structure has none of the paradoxes of QM.

So how are we supposed to think intuitively about the CWL theory? In what follows I try and give some feeling for this.

\subsection{Sums over Correlated Rings}
 \label{sec:CWL-ringst}

Let's start again from the generating functionals in (\ref{DbbQ}) and (\ref{DbbQ2}). It is useful to give a diagrammatic interpretation of these. To do this we introduce a perturbative "graviton expansion" about a background field metric - to be specific let us choose here a background flat space (which is what we will have for earth-based experiments). We can then write
\begin{equation}
\tilde{\mathfrak{g}}^{\mu \nu}(x) = \eta^{\mu \nu} + \lambda h^{\mu \nu}(x)
 \label{h-exp}
\end{equation}
where $\lambda^2 = 16\pi G$, and expand the gravitational Lagrangian $L_G$ in powers of $\lambda h^{\mu \nu}(x)$ in the usual way \cite{donoghue94,barvinsky,GtH74,duff}.

\begin{figure}
\vspace{-0.5cm}
\includegraphics[width=\columnwidth]{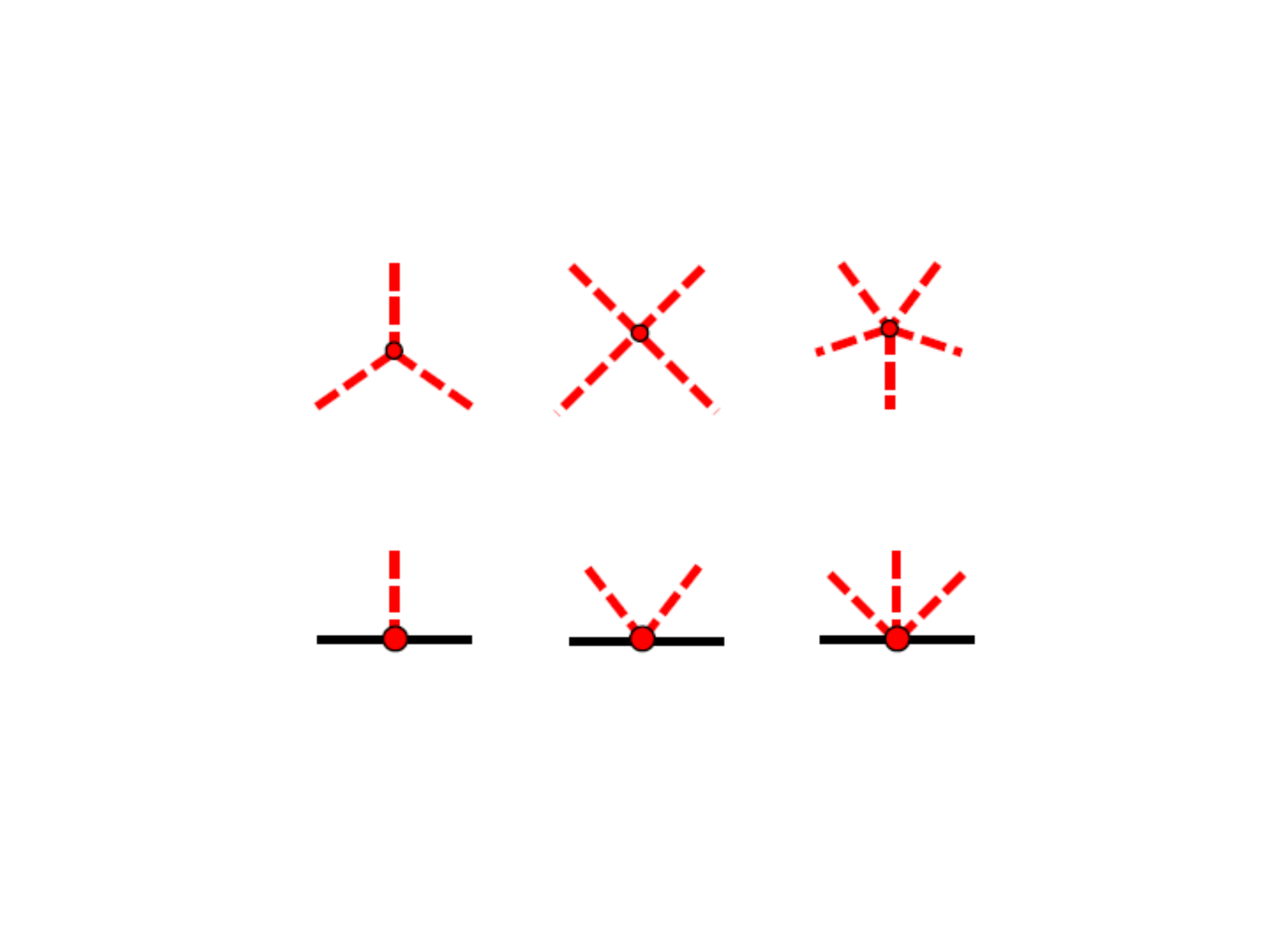}
 \vspace{-1.9cm}
\caption{Graviton graphs involved in perturbative expansions in quantum gravity. The upper graphs are 3rd, 4th, and 5th-order gravition self-interaction vertices; the lower set are graviton-matter interactions.  }
 \label{fig:NJP-F7}
\end{figure}

To see how this works, let us consider first the case of the relativistic particle; the action for this particle was given in (\ref{freeP}) above. To integrate over the metric fluctuations around the background flat space, we write the gravitational Lagrangian as
\begin{equation}
L_G = L_o - \int d^4x \; U(h^{\mu \nu})
 \label{L-G}
\end{equation}
where $L_o$ describes free gravitons and $U(h^{\mu \nu}) \sim O(\lambda)$ describes inter-graviton couplings - in what follows we do not need to know the precise form of these terms, but only their diagrammatic representation - see Fig. \ref{fig:NJP-F7}.


\begin{figure}
\vspace{0.8cm}
\includegraphics[width=\columnwidth]{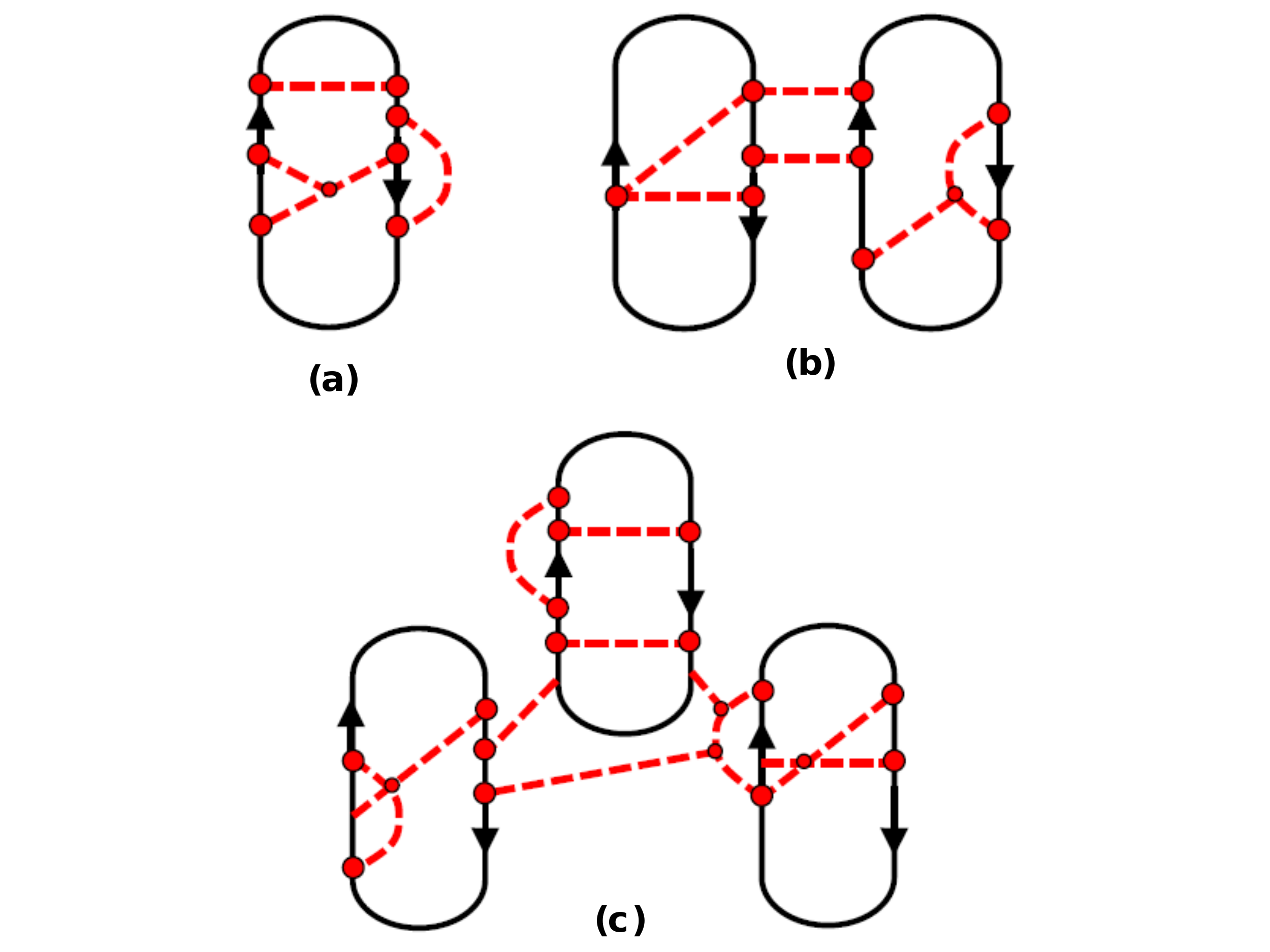}
 \vspace{-0.3cm}
\caption{Diagrams generated in the ring sum $Q[j]$ when we insert the gravitational correlators from eqtn. (\ref{QGrav}). In (a) is shown the conventional QM form $Z[j]$, with internal graviton insertions. In (b) and (c) we see typical graphs for $Q_2[j]$ and $Q_3[j]$, with gravitons mediating correlations between different rings, as well as internal ring 'self-interactions'. }
 \label{fig:NJP-F8}
\end{figure}


Now in conventional quantum gravity, for which the generating functional ${\cal Z}[q]$ was given in (\ref{Z-part}) above, integrating out the metric generates, in a low-energy effective theory, diagrams of the type shown in Fig. \ref{fig:NJP-F8}(a), which renormalize the particle dynamics - diagrams like this follow from eqtn. (\ref{QGrav}). It is crucial to note again that we are dealing with an effective low-energy theory - no integrations up to the Planck energy are involved.

How all this is done is partly a matter of choice - there are various schemes that one can use, including the effective action approach of Barvinsky and Vilkovisky, and/or the approach of Donoghue et al. Although the details here are quite lengthy and technical, and are discussed elsewhere, the basic idea is not so different from that used in regularization schemes in QFT - very high-energy processes are simply absorbed into effective couplings.  The key difference with standard QFT is of course is that we are not necessarily doing the calculation on a flat spacetime background.

One should be aware here of a limitation of such effective theories. All calculations of quantum corrections to a given process refer to a {\it single background field}. Even if this field is curved, it is still the case that all diagrams are referred back to this one spacetime. Thus the causal structure is identical for all the diagrams. This emphasizes the point already made above, that we severely restrict the functional integration over $g^{\mu \nu}(x)$.

Consider now what happens when we add in all the higher correlators $\kappa_n[ \{ q_k \}]$. We have to now deal with a sum over rings. The diagrams generated by eqtn. (\ref{kappa-n}) are shown in Figs. \ref{fig:NJP-F8}(b), \ref{fig:NJP-F8}(c). One should note the way in which the generalized equivalence principle appears here in the diagrams - all lines couple in the same way to each other, whether the coupling refer to a single path in the path integral, or to a coupling between different paths.

The generalization of this scheme to a quantized field is fairly obvious. A nice example is provided by the free scalar field $\phi(x)$, with action (here $\nabla$ denotes a covariant derivative):
\begin{equation}
S[g, \phi] \;=\; {1 \over 2} \int d^4x g^{1/2}\;[g^{\mu \nu} \nabla_{\nu}\phi \nabla_{\mu}\phi - m^2\phi^2]
 \label{S-Phi}
\end{equation}
In conventional QFT the generating functional ${\cal Z}[J]$ for this theory is then
\begin{eqnarray}
{\cal Z}[J] &=&  \int {\cal D}\tilde{\mathfrak{g}}^{\mu \nu} \int {\cal D}\phi_k  \nonumber \\
&\times&  e^{i S_G/\hbar} \; \Delta[\tilde{\mathfrak{g}}^{\mu \nu}] \;  e^{{i \over \hbar}  (S[g, \phi] + \int d^4x \;J(x) \phi(x))  } \;\;\;\;\;\;
 \label{Z-phi}
\end{eqnarray}
in accordance with (\ref{QGrav}) above. On the other hand in CWL theory, the ring sum now becomes:
\begin{eqnarray}
{\cal Q}[J] &=& \sum_{n=1}^{\infty} \prod_{k=1}^n {1 \over n!} \int^{\bf \prime \prime} {\cal D}\tilde{\mathfrak{g}}^{\mu \nu} \int {\cal D}\phi_k  \nonumber \\
&\times&  e^{i S_G/\hbar} \; \Delta[\tilde{\mathfrak{g}}^{\mu \nu}] \;  e^{{i \over n\hbar} \sum_k (S[g, \phi_k] + \int d^4x \;J(x) \phi_k(x))  } \;\;\;\;\;\;
 \label{QG-gen}
\end{eqnarray}

As with the single particle problem, the metric density $\tilde{\mathfrak{g}}^{\mu \nu}(x)$ acts on the different configurations $\phi_k(x)$ via its presence in the action (\ref{S-Phi}), and correlates them via $\kappa_n$. The diagrammatic structure is the same as for the single particle.

It is important at this stage to ask whether the theory defined so far is internally consistent. As with any field theory, one can apply various consistency tests to the CWL theory. These include the calculation of Ward identities, starting from generalized Schwinger-Dyson eqns. for the higher correlators, as well as checking unitarity in the various vertex parts. So far, all of these tests can be shown to be satisfied \cite{GR-Psi-1}.

\subsection{Correlated Rings Sums for higher propagators}
 \label{sec:CWL-GF}

How do we now extract information about real physical quantities from the formal apparatus of these generating functionals? It turns out that there is a subtlety here, coming from the fundamentally non-linear nature of the theory. Field correlators can be defined in a way very much like that in conventional QFT, by functional differentiation of ${\cal Q}[j]$ (or ${\cal Q}[J]$ for a field theory). However if we want to define propagators for field excitations or for particles, describing probability amplitudes for some process, we must do things differently. Let's just quickly look at how this is done formally, and then focus on more detail on how to understand the results intuitively (both diagrammatically and using an example).

\subsubsection{Structure of Correlators}
 \label{sec:CWL-GF-form}

As we noted earlier, we have set up CWL theory here in terms of the generating functional ${\cal Q}[J]$ so that we may easily determine correlation functions by functional differentiation; now it is time to see how this is done.

Again, let us use the the relativistic particle as an example. In conventional QFT one functionally differentiates (\ref{Z-part}) to get the correlators
\begin{equation}
G_n^{\sigma_1,..\sigma_n}(s_1,..s_n) \;=\;
  \left( {-i \over \hbar} \right)^n  \lim_{ j(s) \rightarrow 0} \left[ { \delta^n {\cal Z}[j] \over \delta j(s_1 \sigma_1) ..  \delta j(s_n \sigma_n)} \right]
 \label{Gn-QFT}
\end{equation}
where the $\sigma_j = \pm$ are labels indicating upon which one of the forward/backward segments of the single Kadanoff-Baym ring the currents $j(s_j, \sigma_j)$ intervene.

On the other hand if we functionally differentiate ${\cal Q}[j]$, we have
\begin{equation}
{\cal G}_n^{\sigma_1,..\sigma_n}(s_1,..s_n) \;=\;
  \left( {-i \over \hbar} \right)^n  \lim_{ j(s) \rightarrow 0} \left[ { \delta^n {\cal Q}[j] \over \delta j(s_1 \sigma_1) ..  \delta j(s_n \sigma_n)} \right]
 \label{Gn-NR2}
\end{equation}
where now the external currents $j(s_j, \sigma_j)$, labelled by the $\sigma_j = \pm$, can intervene on any one of the rings in the ring sum. This leads to a very different structure from conventional QFT.

\begin{figure}
\vspace{0.8cm}
\includegraphics[width=\columnwidth]{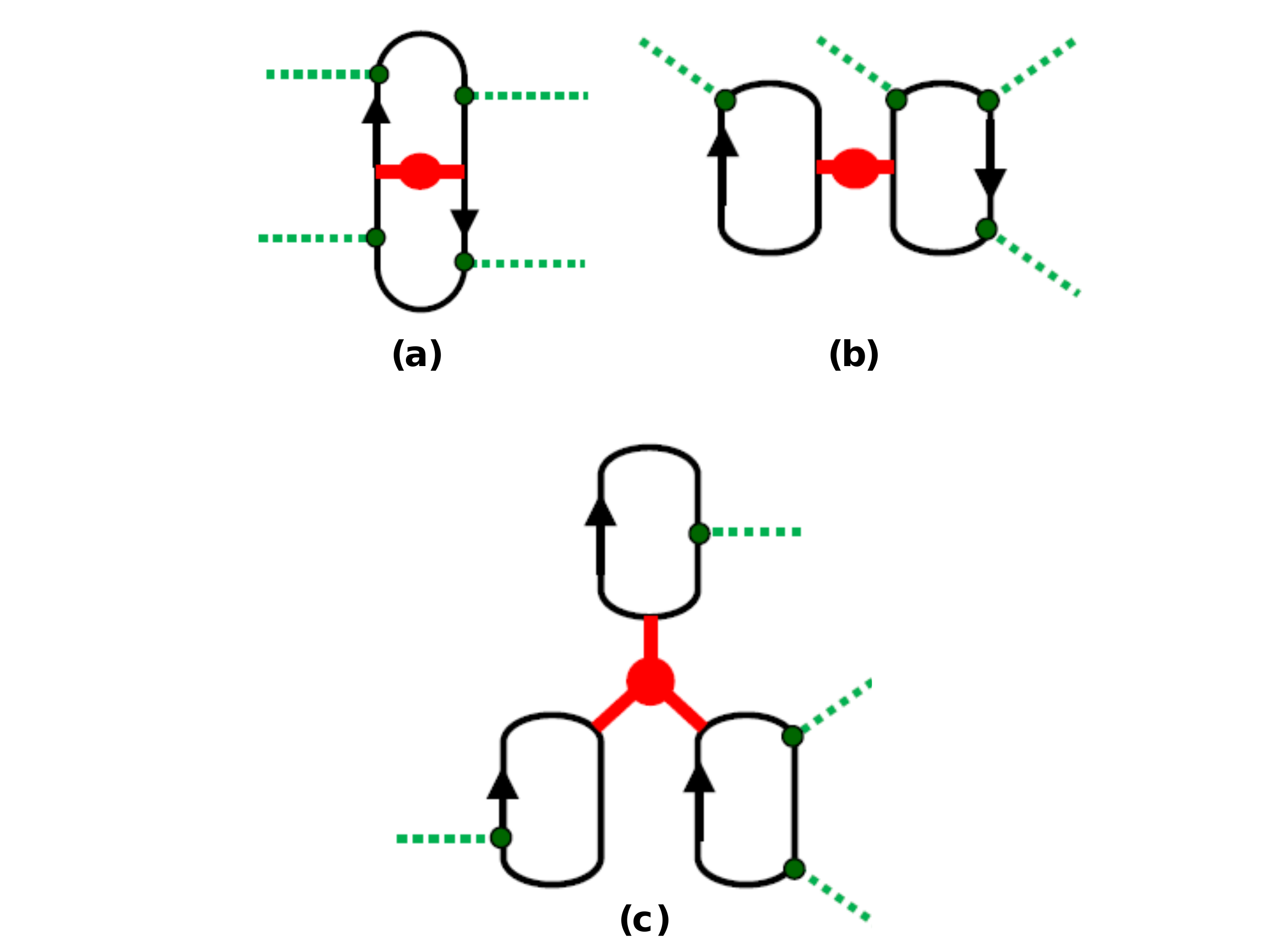}
 \vspace{-0.3cm}
\caption{Some of the graphs contributing to $K_4(\tau_1, ...\tau_4)$, calculated from eqtn. (\ref{CWL-G4}). The external hatched lines are insertions of the external current $j(\tau)$; these may be inserted in all possible ways into the rings. In (a), (b), and (c) we see the contributions involving $\kappa_1, \kappa_2$, and $\kappa_3$ respectively. Thus (a) is showing the conventional QFT result, whereas (b) is showing the first correction to it from 2-ring graphs}
 \label{fig:NJP-F9}
\end{figure}

\begin{widetext}

To see this in a transparent way, without all the functional integrations over spacetime metrics and Faddeev-Popov determinants, let's expand out these results in terms of the correlators $\kappa_n$. Then for conventional QFT one just gets
\begin{equation}
G_n^{\sigma_1,..\sigma_n}(s_1,..s_n) \;=\;  \int {\cal D}q(\tau)\; e^{{i \over \hbar} S_o[q]} \; \kappa_1[q] \; \prod_{j=1}^n   q(s_j, \sigma_j)
 \label{QFT-Gn2}
\end{equation}
whereas for the CWL theory, substituting (\ref{Gn-NR2}) into (\ref{DbbQ}), we get
\begin{equation}
{\cal G}_n^{\sigma_1,..\sigma_n}(s_1,..s_n) \;=\; \sum_{r=1}^{\infty} \prod_{\alpha = 1}^r \int {\cal D}q^{\alpha}(\tau)\; e^{{i \over r \hbar} \sum_{\alpha} S_o[q^{\alpha}]} \; \kappa_r(\{ q^{\alpha} \}) \; \prod_{j=1}^n \left( \sum_{\alpha = 1}^r q^{\alpha}(s_j, \sigma_j)\right)
 \label{CWL-Gn2}
\end{equation}
so that the $n$-point correlation function now involves insertions of $n$ different currents at all possible combinations of points on the multi-ring diagrams (see Fig. \ref{fig:NJP-F9}).

This structure becomes clearer if we take a specific example, and show the results graphically. Consider, eg., the 4-point correlator ${\cal G}_4^{\sigma_1,..\sigma_4}(s_1,..s_4)$, in which an external current intervenes at 4 different times in the functional differentiation.
Expanding out (\ref{CWL-Gn2}) for ${\cal G}_4$, we find:
\begin{eqnarray}
{\cal G}_4^{\sigma_1,..\sigma_4}(s_1,..s_4) \;&=&\; \oint {\cal D}q \; e^{{i \over \hbar} S[q]} \; \kappa_1[q] \; \prod_{j=1}^4 q(s_j, \sigma_j) \nonumber \\
&+& {1 \over 2} \; \oint {\cal D}q \oint {\cal D}q' \; e^{{i \over 2\hbar} (S_o[q] + S_o[q'])} \; \kappa_2[q,q'] \; \prod_{j=1}^4 \;[q(s_j, \sigma_j) + q'(s_j, \sigma_j)]  \;\;\;\;+\;\;\; etc....\;\;\;\;\;\;\;
 \label{CWL-G4}
\end{eqnarray}

In ordinary QFT we only have the first term on the right hand side; it defines $G_4^{\sigma_1,..\sigma_4}(s_1,..s_4)$, involves only a single path $q(s)$, and is measuring the correlation between 4 different positions of the particle on this path (and in conventional QFT one determines the expectation value for the measurement of this 4-point correlation from this first term). In contrast the second term on the right hand side of (\ref{CWL-G4}), ie., the lowest correction to the conventional QFT result, involves pairs of paths $q, q'$ for the {\it same particle}, and has current insertions on both of these paths - with 4 current insertions we can have all 4 insertions on one path, or 3 on one path and 1 on the other, and so on; we must sum external current insertions over all the topologically different arrangements. All of this is illustrated in Fig. \ref{fig:NJP-F9}.

The above results are purely formal. If we now insert into these results the specific form of the $\kappa_n$ that arises in the CWL theory, we get
\begin{equation}
{\cal G}_n^{\sigma_1,..\sigma_n}(s_1,..s_n) \;=\; \sum_{r=1}^{\infty}
{1 \over r!} \int^{\bf \prime \prime} {\cal D}g^{\mu\nu} e^{{i \over \hbar} S_G[g^{\mu \nu}]} \Delta[g^{\mu \nu}] \prod_{\alpha = 1}^r \int {\cal D}q^{\alpha}(\tau)\; e^{{i \over r\hbar} \sum_{\alpha} S_o[q^{\alpha}]}  \; \prod_{j=1}^n \left( \sum_{\alpha = 1}^r q^{\alpha}(s_j, \sigma_j)\right)
 \label{CWL-Gn2'}
\end{equation}
and if we remove the sum $\sum_r$ and the product $\prod_{\alpha}$ in this expression, we get back the form for the correlator of a particle in ordinary quantum gravity. As we might expect, this result is clearly highly non-linear, but its properties are not immediately obvious. The main problem here is that in a non-linear theory of this kind, it is not clear how to connect the correlators with physical properties of the system.

\vspace{2mm}

\end{widetext}

\subsubsection{Structure of Propagators}
 \label{sec:CWL-prop-form}

A much more physical question to ask is - how do how particles or field excitations propagate in the CWL theory? We will look at this here in two ways. First, we look at perturbative expansions in the graviton variables $h^{\mu\nu}(x)$ given in (\ref{h-exp}), applied to the correlators in the theory, and see how these expansions are interpreted diagrammatically. Then we look at the example of a single particle, again in a perturbative expansion, to see what quantitative results emerge.

In a graviton expansion, all the propagators are written in the form of a functional integral over the gravitons. This lends itself easily to a diagrammatic interpretation. Consider, eg., what form the density matrix propagator ${\cal K}_{2,2';1,1'}$ takes for the relativistic particle. Then we have


\begin{eqnarray}
{\cal K}_{2,2';1,1'} &=& lim_{\hat{h} = 0} \;\{ e^{{i \over 2 \hbar} (\delta_{\hat{h}} |\hat{D}| \delta_{\hat{h}'})  } \nonumber \\
&\times& e^{{-i \over \hbar} \int U(\hat{h})}\;
{\cal K}_{2,1}[\hat{h}(x)]\; {\cal K}_{1',2'}[\hat{h}(x')] \}
 \label{G4mss}
\end{eqnarray}
where $\hat{h}(x) = h^{\mu \nu}(x)$ is the graviton field, the graviton propagator $\hat{D}(x) = D^{\mu \nu \lambda \rho}(x)$, and we define
\begin{equation}
(\delta_{\hat{h}} |\hat{D}| \delta_{\hat{h}'}) \;=\; \int d^4x d^4x' {\delta \over \delta \hat{h}(x)} \hat{D}(x,x') {\delta \over \delta \hat{h}(x')}
 \label{hDh}
\end{equation}
in which
\begin{equation}
{\cal K}_{2,1}[\hat{h}(x)]  \;=\; \sum_{n=1}^{\infty} \prod_{j=1}^n \;{1 \over n!} \; K_j[2,1|\hat{h}(x)]
 \label{G1-sum}
\end{equation}
and where $K_j(2,1|\hat{h}(x))$ is just the conventional QM particle propagator in the weak field $\lambda \hat{h}(x))$.


\begin{figure}
\vspace{-0.2cm}
\includegraphics[width=\columnwidth]{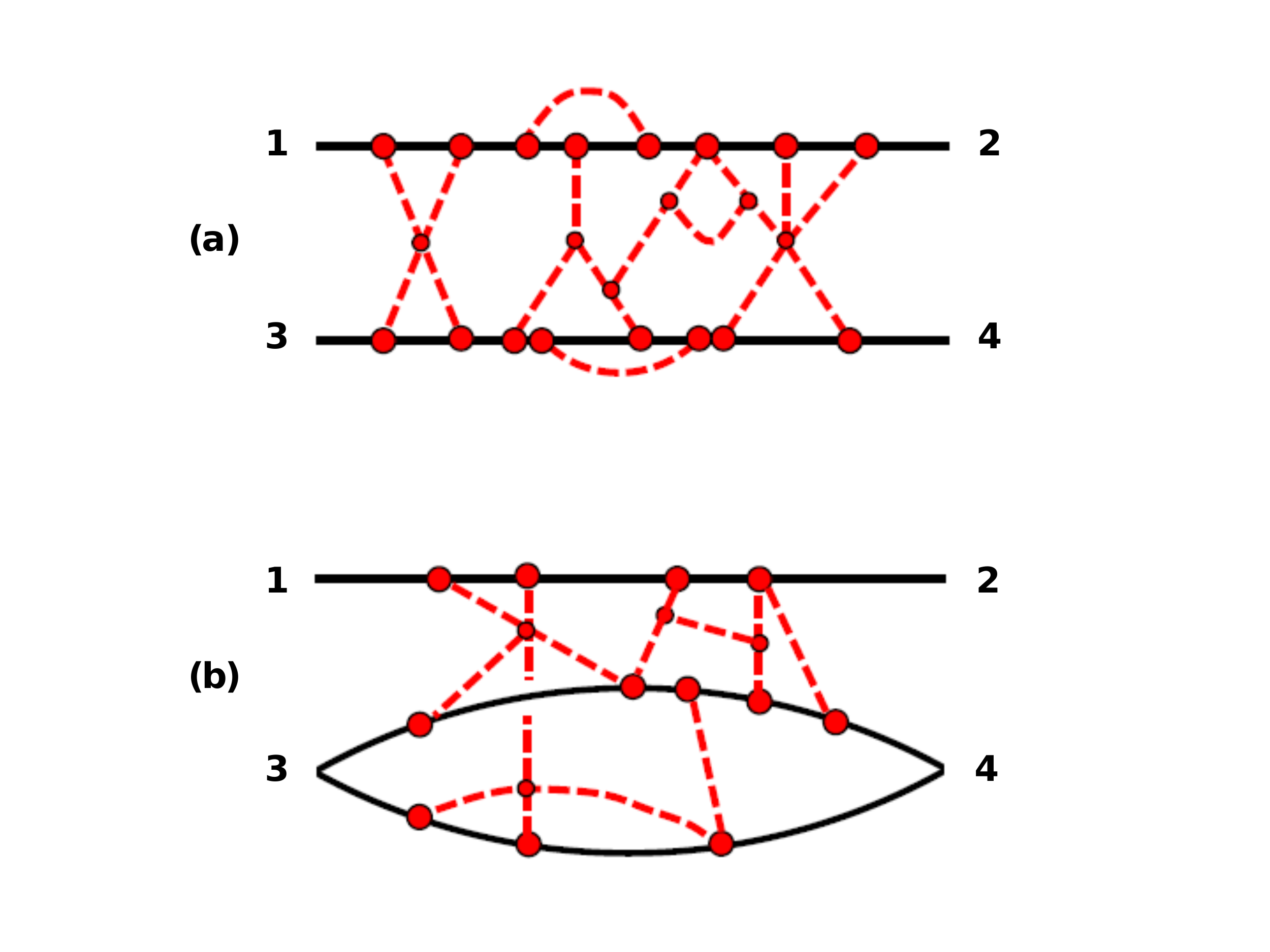}
 \vspace{-0.8cm}
\caption{Two typical graphs for the propagator $K_4(\tau_1, ...\tau_4)$ for 2 particles, with multiple graviton insertions. In (a) we have a graph which exists in conventional quantum gravity, but (b) involves internal gravitational correlations between different paths of one of the particles. }
 \label{fig:NJP-F10}
\end{figure}


This generates diagrams of the type shown in Fig. \ref{fig:NJP-F10}. In Fig. \ref{fig:NJP-F10}(a) we see the kind of graph that would be generated in conventional quantum gravity for ${\cal K}_4^{\sigma_1,..\sigma_4}(s_1,..s_4)$. However in Fig. \ref{fig:NJP-F10}(b) we see a graph in which one of the two internal particle lines is doubled - this graph includes both correlations between the two matter lines, and correlations between the 2 particle lines, coming from $\kappa_2$.

We see from eqtns. (\ref{G4mss})-(\ref{G1-sum}), and from Fig. \ref{fig:NJP-F10} that, as we might expect, propagators like ${\cal K}_{2,2';1,1'}$ depend on the full ring sum form for the lower propagator ${\cal K}_2(1,1')$.


\begin{figure}
\vspace{-0.2cm}
\includegraphics[width=\columnwidth]{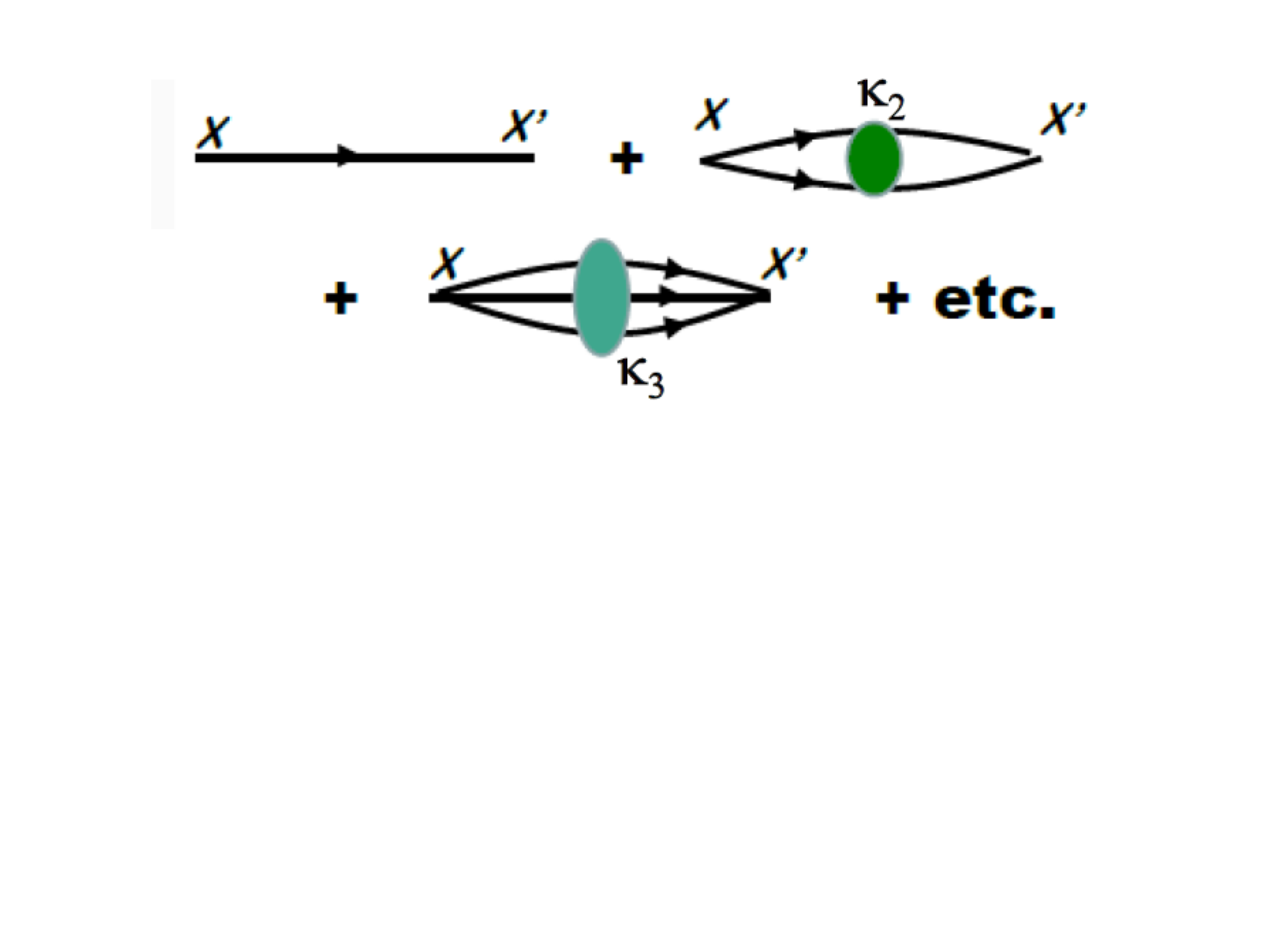}
 \vspace{-4.1cm}
\caption{Expansion of the propagator ${\cal K}_2(1,1')$ in powers of the correlators $\kappa_n$. }
 \label{fig:NJP-F11}
\end{figure}


However, as we might also expect, the higher propagators also feed back in the usual way on the lower ones. A systematic discussion of this is dobne using Schwinger-Dyson equations for the CWL theory \cite{GR-Psi-1}, but one can understand this intuitively by looking at how it works for the ordinary propagator ${\cal k}(1,1')$.
Fig. \ref{fig:NJP-F11} shows the diagrammatic development for ${\cal K}(1,1')$; formally we simply write ${\cal K}(1,1') = K_o(1,1') + \Delta {\cal K}(1,1')$, where
\begin{equation}
\Delta {\cal K}(2,1) \;\;=\;\;  \int_1^2 {\cal D}q \int_1^2 {\cal D}q' \; \kappa_2[q,q'] e^{{i \over 2\hbar}(S[q] + S[q'])} \; + \; ...
 \label{DG21}
\end{equation}
where $q, q'$ are 4-vectors, and the higher terms on the right-hand side of (\ref{DG21}) involve $\kappa_3[q,q',q'']$ operating between 3 paths, $\kappa_4[q,q',q'',q''']$ operating between 4 paths, and so on. One has a similar set of equations for ${\cal K}_4(1,2;1',2')$, and so on.

\subsubsection{Single Particle at low velocity}
 \label{sec:CWL-GF-singleP}

To gain more intuition for the theory, and appreciate its implications for quite ordinary physical phenomena, it is helpful to take the simplest possible example. We therefore look at the dynamics of a single free particle in the non-relativistic regime, ie., a particle moving in free space with non-relativistic velocity $v \ll c$. We stress that the following calculation is done purely in order to show how the theory works - to compare with experiments on massive bodies, which are composed of many particles, organized in some way in space, we have to do much more sophisticated calculations \cite{GR-Psi-1}.

For a single non-relativistic particle the conventional quantum mechanics is described by the free particle propagator $K_o(x_2,x_1)$, where $x_j = ({\bf r_j}, t_j)$, with $j = 1,2$.  This takes the usual form
\begin{equation}
K_o(x_2,x_1) \;=\;  A_o(2,1) \; \exp iS_c/\hbar
 \label{G-freeP}
\end{equation}
where the prefactor is $A_o(2,1) = [m/2 \pi i\hbar (t_2 - t_1)]^{1/2}$, and the classical action is $S_c = m |{\bf r_2} - {\bf r_1}|^2/2(t_2-t_1)$.

Now consider how to describe a free particle in the CWL theory. We begin by writing the correlator $\kappa_2[q,q']$ in the form
\begin{equation}
\kappa_2[q,q'] = {1 \over 2} \left[e^{i \chi_2[q,q']}\;-\;\delta_{qq'}\right]
 \label{kappa2-chi}
\end{equation}
where the phase $\chi_2[q,q']$ is a (in general complex) functional of the paths $q(\tau), q'(\tau)$; the factor $\delta_{qq'} = \delta(q(s) - q'(s))$ excludes any pair of identical paths in the double path integral in (\ref{DG21}) (because these are already in the bare Green function (\ref{G-freeP})).

We can now calculate the lowest correction $\Delta {\cal K}(2,1)$ to the full propagator ${\cal K}(2,1)$, by doing a perturbative expansion in $\lambda$ for the double path integral in (\ref{DG21}). In the non-relativistic regime $q \rightarrow ({\bf q}, t)$, and defining ${\bf r} = {\bf q} - {\bf q'}$, one finds \cite{GR-Psi-L1,GR-Psi-1} that
\begin{equation}
\chi_2[q,q'] \;=\; \chi_2^N[q,q'] + \delta \chi_2[q,q']
 \label{chiqq'A}
\end{equation}
where the leading term is the "Newtonian phase", given by
\begin{equation}
\chi_2^N[q,q'] \;=\;  \int^t {d\tau \over 4 \pi \hbar} {m^2\lambda^2 \over |{\bf r}(\tau)|}
 \label{chiqq'N}
\end{equation}
and, using standard methods \cite{donoghue94,hamber}, the correction term to this is found to be \cite{GR-Psi-L1,GR-Psi-1}:
\begin{equation}
\Delta \chi_2[q,q'] \;=\;  \int^t {d\tau \over 4 \pi \hbar} {m^2\lambda^2 \over r(\tau)}
 \left[ {R_s \over r(\tau)} - {122 \over 15 \pi} {L_p^2 \over r^2(\tau)} + ...\right] \;\;\;\;\;\;\;
 \label{chiqq'C}
\end{equation}
where $R_s = 2Gm/c^2$ is the Schwarzchild radius for the mass $m$, and $r = |{\bf r}|$.

The leading term (\ref{chiqq'N}), which dominates for $|{\bf r}(\tau)| \gg R_s, L_p$, is just the single graviton term, which in this low-velocity limit gives a Newtonian interaction between the pairs of paths in $\Delta {\cal K}(x_2, x_1)$. The higher correction $\Delta \chi_2[q,q']$ is negligible unless the particle paths are in some way constrained to be within a distance $\sim L_p$ and/or $\sim L_p$ of each other; neither possibility is remotely feasible in any earth-based experiment.

These are not the only corrections to $\Delta {\cal K}(x_2, x_1)$. If we go to correlations between triplets, quadruplets, etc., of paths, we then get Newtonian contributions from these as well. Their analysis is more complicated \cite{GR-Psi-L1,GR-Psi-1}, but we return them briefly below.

What does all this then mean for the propagation of a free particle of mass $m$? Let's consider an arbitrary pair of paths. Since they are attracted to each other by a Newtonian potential, we can characterize this potential by a length scale
\begin{equation}
l_G(m) = \left( {M_p \over m} \right)^3 L_p
 \label{l-G}
\end{equation}
which is the Newtonian gravitational analogue of the Bohr radius for an attractive Coulomb interaction.

There is also an independent energy scale in the system, given by
\begin{equation}
\epsilon_G(m) = G^2m^2/l_G(m) \equiv E_p (m/M_p)^5
 \label{gravE}
\end{equation}
which is the "Newtonian binding energy" for this attractive interaction; here $E_p = M_pc^2$ is again the Planck energy.

Notice that both of these expressions vary extremely rapidly with the particle mass. They play a direct role in the dynamics - paths approaching each other closer than the length $l_G(m)$ will be attracted to each other, and can even bind together, provided there are no other energies acting on the paths that are larger than  $\epsilon_G(m)$.

Thus we arrive at a startling conclusion. There is what we might call a "path bunching" mechanism operating here - for sufficiently large masses, different paths cannot "escape from each other". If the universe were infinite in extent and utterly empty, with space completely flat everywhere, this mechanism would operate for arbitrarily small masses. However it is physically obvious in the case of an electron, where the binding energy $\epsilon_G(m)$ is $\sim 10^{-84}$ eV, and the gravitational binding length is roughly 3.6 million times the Hubble radius, that this bunching mechanism can play no role whatsoever - the binding energy is insignificant compared with any other coupling acting on a real electron.

If we add the higher correlators to the 1-particle calculation just given, all that we find is that the bunching mechanism is enhanced still further: 3-path, 4-path, etc., attractions merely add to what one already finds in this 2-path calculation. Note that when the mass $m$ is significantly less than the crossover mass $m_c$, these corrections are very small. However a proper treatment of the crossover requires that they be included.

Consider however what happens for particle masses in the range $10^{-15}$kg $< m < 10^{-14}$kg. When $m = 10^{-15}$kg, we have $\epsilon_G(m) = 2.6 \times 10^{-9}$eV (or $30 \mu$K in temperature units), and $l_G(m) = 1.67 \times 10^{-13}$m (ie., $~150$ nucleon diameters). On the other hand when $m = 10^{-14}$kg, we have $\epsilon_G(m) = 2.6 \times 10^{-4}$eV (or $3$K in temperature units), and $l_G(m) = 1.67 \times 10^{-16}$m (ie., $1/6$-th of a nucleon diameter). We see that in this range, $\epsilon_G(m)$ makes an extremely rapid passage through laboratory energy scales - we pass from a purely quantum regime, where the paths propagate independently, to a purely classical regime, in which they are bound together. Thus one sees a sudden "collapse" of the quantum correlations, as the bunching mechanism operates; this happens around a "crossover mass" $m_c \sim 10^{-14}$ kg.

Now these numbers are not directly relevant to any real experiment, for two reasons:

(i) in any experiment, or indeed any other real physical process, $\epsilon_G(m)$ will be competing with other energies in the experimental setup that will destabilize the tendency for the paths to bunch. Only if $\epsilon_G(m)$ is larger than these other energies will the path-bunching mechanism operate. Note that amongst these other energies will be the coupling of the particle to other dynamical degrees of freedom, which will cause "environmental decoherence" in the dynamics of the particle.

(ii) more importantly, any real mass in the range $\sim 10^{-15}-10^{-14}$kg will not be in the form of a point particle (unless we deal with an ultra-relativistic particle, in which case the non-relativistic approximation used here is not applicable). In any laboratory set-up such a mass will be spread out, in the form of a solid object, and to evaluate the dynamics we must use a proper interacting $N$-particle theory of this solid object, which incorporates these interactions. This turns out to be rather lengthy, and the final result depends rather crucially on the form of the interparticle interactions \cite{GR-Psi-1}.

A key point, in connection with (i), should be emphasized at this point. The path bunching mechanism here is {\it not a decoherence mechanism}, in any meaningful sense. Thus, path bunching does not involve information loss to, or dephasing by, any environment, via entanglement of the system degrees of freedom to this environment - the bunching mechanism operates for a system in isolation. Even in the absence of the usual decoherence mechanisms, there will of course be gravitational decoherence in the dynamics of the particle, via coupling to background gravitons  - but even when kinematically possible, this is an extremely weak process, and has nothing to do with the bunching mechanism discussed here.

One can of course define a "bunching" timescale $\Delta t_G = \hbar/\epsilon_G(m)$ associated with the binding energy scale $\epsilon_G(m)$; but this timescale simply characterizes a process occurring for a closed system, and has nothing to do with environmental decoherence.

It is also useful to compare what we get here with the "Diosi-Penrose" formula noted in section \ref{sec:mod-Sch} (which is, on the other hand, supposed to characterize a decoherence mechanism). Suppose we expand the exponential in $\kappa_2^{(1)}[q,q']$ in eqtn. (\ref{kappa2-chi}), in a power series; the first term by itself gives a phase correction
\begin{equation}
\Delta \chi[q,q'] \;=\; {\lambda^2 m^2 \over 4\pi \hbar} \int d\tau  {1 \over |{\bf q}(\tau) - {\bf q'}(\tau)|}
 \label{phiC}
\end{equation}
which we compare with the expressions of Penrose \cite{penrose96} and Diosi \cite{diosi90} written in the form of a phase correction
\begin{equation}
\Delta \phi \;=\; {\lambda^2 m^2 \over 4 \pi \hbar} \int d\tau \int d^3q d^3q' {\rho({\bf q}) \rho({\bf q'}) \over |{\bf q}(\tau) - {\bf q'}(\tau)|}
 \label{phiC2}
\end{equation}
where $\rho({\bf q})$ is a density distribution for the particle, introduced here by hand. However there are 2 problems with this manouevre. First, as we already noted in section \ref{sec:mod-Sch}, in any quantum field theory the density distribution $\rho({\bf q})$ should come out of the theory, rather than being inserted {\it ad hoc}, and it will be dependent on the effective UV cutoff. Second, it is of course incorrect to expand the exponential in $\kappa_2^{(1)}[q,q']$ in a power series, since each term diverges, and only the total exponent is meaningful (thus, even if the first problem could be dealt with, the Penrose-Diosi formula would give a radically different result from the CWL formula).

Nevertheless, it is interesting that the inter-path correlations in the CWL theory do lead to phase corrections of the kind discussed qualitatively by Penrose \cite{penrose96}, even if they may be quantitatively very different. But let us note again that the "path bunching" mechanism involves no dissipation or decoherence.

\subsubsection{Classical Limit}
 \label{sec:CWL-class}

Discussions of the classical limit are subtle in any quantum theory, because this limit is singular; the pitfalls have become more apparent in recent years, with detailed study of the asymptotic properties of expansions in powers of $\hbar$  \cite{berry-cl}. In quantum field theory texts it is typically argued that one can extract the classical behaviour from a loop expansion, with the classical behaviour encoded in the sum of all tree graphs - this result goes back to work of Nambu \cite{nambu66}. However, this is actually incorrect \cite{holstein04}; in the presence of massless fields, infra-red singular terms from loop diagrams also contribute to the classical theory.

There is still no systematic understanding of these contributions, and they are particularly important in quantum gravity - indeed, the contribution of loop diagrams to classical GR has been known for a long time \cite{donoghue94,deser70,iwasaki}. This confusion has shown itself in recent years in different calculations of the corrections to Newton's law in quantum gravity \cite{hamber}, where the quantum and classical contributions have to be disentangled in any diagrammatic expansion. Note that such expansions should be distinguished from the diagrammatic methods that have been used to do weak-field expansions (up to 7th post-Newtonian) around Newtonian theory, which also have a long history \cite{das,damour,rothstein,detweiler}, and have been used to analyze, amongst other things, gravitational inspiraling, tidal interactions, gravitational self-interactions, caustic echoes around black holes, compact binary and triplet systems, and various other features of black hole dynamics. These latter effects are of course utterly negligible for the laboratory-scale masses we have been referring to in this paper.

It is clear that in the regime where quantum effects are still important in the CWL theory (ie., for small masses), sorting out the various contributions, and deciding which are classical and which are quantum, will be even more complex than it is for conventional Quantum Gravity. Luckily we do not have to do this for massive bodies - once the mass exceeds the crossover mass, the theory becomes entirely classical, since the path bunching mechanism simply forces all paths together. At this point one can develop a diagrammatic calculus along the same lines done for the classical theory - this is of course quite a lengthy task, but one expects that it will reproduce the classical results.

\subsection{Wave-functions, Measurements, and all that}
 \label{sec:CWL-QM-mmt}

Finally, a brief word about how one discusses quantum measurements in the CWL theory. These appear as a specific kind of physical process, albeit a rather involved one. A proper discussion of this topic needs a whole paper \cite{GR-Psi-1}; here we simply explain very briefly how measurements fit into the CWL picture, and how the very rapid transition from quantum-mechanical to classical dynamics occurs in the theory.

\subsubsection{Measurements in ordinary QM}
 \label{sec:QM-mmt}

Recall that in conventional QM, one is supposed to describe a quantum system by a state-vector, and measurements are supposed to be described by the interaction of the quantum system with some classical measuring apparatus, which causes a wave-function projection or "wave function collapse". Thus the measuring system and the measurement operation are considered to be irreducible; indeed the measuring system is considered to be an {\it extraneous non-quantum agent}. The world is thereby divided between "quantum systems", and "observers". To anyone not used to this, the whole idea seems hopelessly anthropocentric, indeed rather mediaeval - and it has of course been heavily criticized, by Bell \cite{bell} and many others\cite{laloe}.

\begin{figure}
\vspace{0.1cm}
\includegraphics[width=\columnwidth]{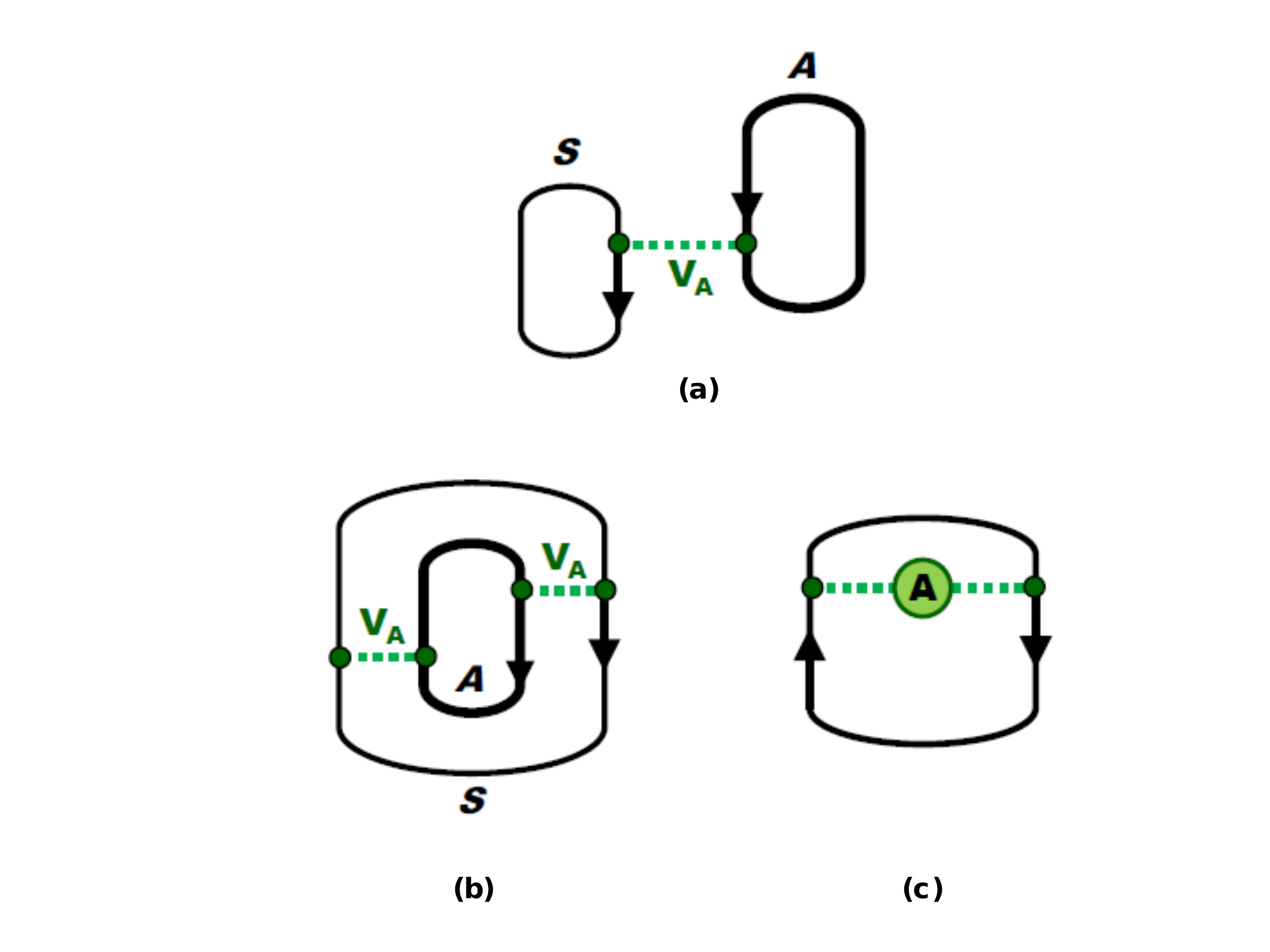}
 \vspace{-0.5cm}
\caption{Interaction between a system ${\cal S}$ and an apparatus ${\cal A}$ in conventional QM. In (a) we see the basic interaction $V_A$ between ${\cal S}$ and ${\cal A}$:(b) shows ${\cal Z}_{SA}$ in 2nd-order perturbation theory in $V_A$. If we integrate out the apparatus ${\cal A}$ we obtain a "reduced" $\bar{\cal Z}_S$ for ${\cal S}$, shown in (c). }
 \label{fig:NJP-F13}
\end{figure}

Here we will instead discuss standard QM using path integrals, avoiding the use of operators and wave-functions. Consider first the interaction of two different systems in QM (Fig. \ref{fig:NJP-F13}). This is described by single ring diagrams, and to make things simple, we begin by ignoring gravity completely. We call these 2 systems ${\cal S}$ (for "system") and ${\cal A}$ (for "apparatus"), and their generating functional is
\begin{equation}
{\cal Z}_o[j,J] \;\;=\;\; {1 \over {\cal N}_o} \int {\cal D}q {\cal D}\phi \; e^{{i \over \hbar}(S[q,\phi] + \int (jq + J\phi))}
 \label{S-S+M}
\end{equation}
where $S[q,\phi]$ describes the coupled system, and the system and apparatus coordinates $q$ and $\phi$ couple to external sources $j$ and $J$ respectively. We can also couple the apparatus (and the system, if we wish) to an "environment" ${\cal E}$ with coordinate field $\chi(x)$ and source $I(x)$, for a total generating functional
\begin{equation}
{\cal Z}[j,J,I] \;\;=\;\; {{\cal Z}_o \over {\cal N}'} \; e^{{i \over \hbar}(S_{int}[q,\phi,\chi] + S_E[\chi] + \int I\chi )}
 \label{S-S+M+E}
\end{equation}
where $S_{int}[q,\phi,\chi]$ describes the interaction between all 3 systems, and ${\cal N}'$ is the correction to the vacuum-to-vacuum amplitude.

Now at this point we have a choice. We can:

(i) simply follow the correlations as they spread ever further throughout the universe, without ever invoking any kind of projection or measurement. This is perhaps the natural thing to do in this framework - it is a path integral version of the Everett "relative state" idea, in its original form \cite{everett57}. Alternatively we can:

(ii) functionally average over either the system or the environment, to project out the dynamics of the apparatus. This choice singles out measurements and measurement interactions for special treatment. In order for this manoeuvre to yield probabilities, according to the Born rule, the functional averaging needs to incorporate time asymmetry \cite{harveyB,GR-Psi-1}.


The key defining characteristic of a measurement is the existence of an interaction $V_A$ between ${\cal S}$ and ${\cal A}$ which correlates the paths of the 2 systems. Typically one demands that the paths of ${\cal A}$ "after" the measurement interaction $V_A$ are correlated with certain sets of paths of ${\cal S}$ existing "before" the interaction. The assumed time-asymmetry is typically imposed using boundary conditions - we can do this by fixing an "initial distribution" over paths at some initial time $t_1$ (or on some initial Cauchy surface, cutting through all the paths), in the form of a density function $\rho_{in}(q,q';\phi,\phi')$. This is a kind of "state preparation"; we throw away all information about the paths "before" this initial state, and select out certain paths over others in our starting point. If this state preparation also involves an averaging over ${\cal A}$ and/or ${\cal E}$, we will be left with an initial "reduced density function" $\bar{\rho}_{in}(q,q')$.

The subsequent dynamics of $\rho(q,q';\phi,\phi')$ or $\bar{\rho}_{in}(q,q')$ is then determined by the correlator $G_4$, or suitable averages over it. Thus, eg., we have
\begin{equation}
\bar{\rho}(2,2') = \int d1 \int d1' \bar{K}(2,2';1,1') \bar{\rho}(1,1')
 \label{rKr}
\end{equation}
where $\bar{\rho}(1,1')$ means $\bar{\rho}(q_1, q_1^{\prime}; t_1)$, and where the worldline pair propagator $\bar{K}(2,2';1,1')$ (found from the 4-point correlator $G_4$ after averaging over ${\cal A}$ and/or ${\cal E}$) is given by the path integral
\begin{eqnarray}
\bar{K}(2,2';1,1') &=& \int_1^2 {\cal D}q(t) \int_{1'}^{2'} {\cal D}q'(t) \nonumber \\
&& \;\;\;\;\;\; \times \; {\cal F}[q,q'] \; e^{{i \over \hbar}(S_o[q] - S_o[q'])}
 \label{K-F}
\end{eqnarray}
with the end points at the arguments of the two different density matrices; here ${\cal F}[q,q']$ is the usual Feynman influence functional \cite{feynmanV63}.

In orthodox QM one calculates the results of measurements of a particular physical quantity on a specific system ${\cal S}$ using quantities defined for ${\cal S}$ only, without referring in any way to the interaction $V_A$, or to ${\cal A}$. To do this here, one must relate the interaction $V_A$ to the quantity $M_V$ that one is trying to define for the system ${\cal S}$, in the form
\begin{equation}
 \langle M \rangle = \int dx dx' \; M_V(x,x') \; \bar{\rho}(x',x)
 \label{rhoVM}
\end{equation}
To make this connection is in general a lengthy business, even for quite simple weak interactions (where one can do perturbation theory in $V_A$ to derive $M_V(x,x')$ in terms of $V_A$); we have no space here to go through this (see ref. \cite{GR-Psi-1}).

To summarize - in orthodox QM, to discuss measurements we must throw away information - about, eg., the interaction $V_A$ - and we must introduce a distinction between past and future (and throw away or average over information about the future).  Notice that at no point are we forced to introduce wave-functions (although in practise they can be very useful!); ordinary QM can be analysed entirely in terms of the density matrices.

\subsubsection{Measurements in CWL theory}
 \label{sec:CWL-mmt}

\begin{figure}
\vspace{-0.7cm}
\includegraphics[width=\columnwidth]{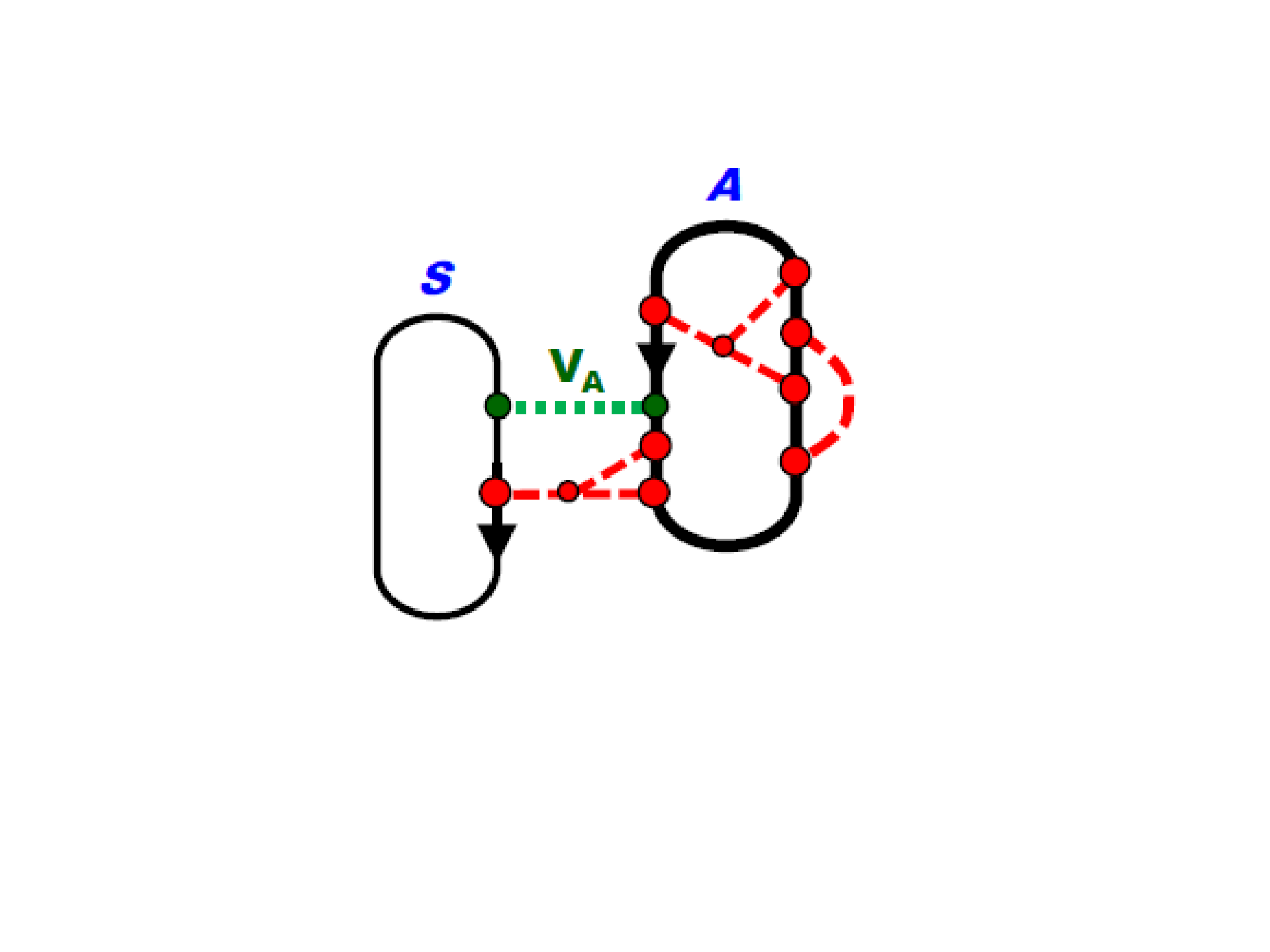}
 \vspace{-2.4cm}
\caption{Interaction between a microscopic system ${\cal S}$ and a "macroscopic system" ${\cal A}$, via an interaction which correlates their paths.  }
 \label{fig:NJP-F14}
\end{figure}

Now let us add back in the CWL correlations $\kappa_n$ coming from gravity. At this point things become quite interesting (Fig. \ref{fig:NJP-F14}). Suppose that ${\cal S}$ is microscopic - so that the gravitational correlations have a negligible effect - but that ${\cal A}$ is large, with a mass $> 10^{-14}$kg, so that "path bunching" forces its real space dynamics to be essentially classical. Suppose also that the interaction between ${\cal S}$ and ${\cal A}$ is actually between the coordinates $q,\partial/\partial{q}$  of ${\cal S}$ and a "collective coordinate" $Q$, which we will assume to be associated with a mass $M$ (eg., the centre of mass of some part of the apparatus) along with its derivative $\partial/\partial{Q}$. We have in effect separated $Q$ from the rest of the field variables $\phi(x), \partial_{\mu}\phi$ describing ${\cal A}$. A simple example of such an interaction would be
\begin{equation}
V_A(q,Q) = -i \hbar V(q,\partial/\partial{q};t)\; {\partial \over\partial{Q}}
 \label{V-A-ex}
\end{equation}
where the function $V(q,\partial/\partial{q};t)$ switches on and off over a time period $T_{mmt}$ - we assume that this interaction is designed so that it does not appreciably excite the other degrees of freedom of ${\cal A}$ (which can be treated as part of the environment ${\cal E}$). Thus $V_A(q,Q)$ is switched on and off slowly - we can treat it non-relativistically, and we will assume that the other variables pertaining to ${\cal A}$ couple adiabatically to $Q$ and $\partial/\partial{Q}$.

Now $V_A(q,Q)$ is designed to have the effect, over a timescale $T_{mmt}$, of strongly correlating the ${\cal S}$ paths with ${\cal A}$ paths - in such a way that in conventional QM we would say that the position variable $q$ of ${\cal A}$ correlates with the collective variable $Q$ of ${\cal A}$. In path integral language, typical values for $q$ in the paths for ${\cal S}$ would be correlated with typical values for $Q$ in the paths for ${\cal A}$. This is a rather simple form for the measurement interaction - real measurements are typically much more complicated, particularly when one is not measuring position but, eg., momentum (where typically one performs sequences of measurements, as in quantum non-demolition protocols).


\begin{figure}
\vspace{-0.9cm}
\includegraphics[width=\columnwidth]{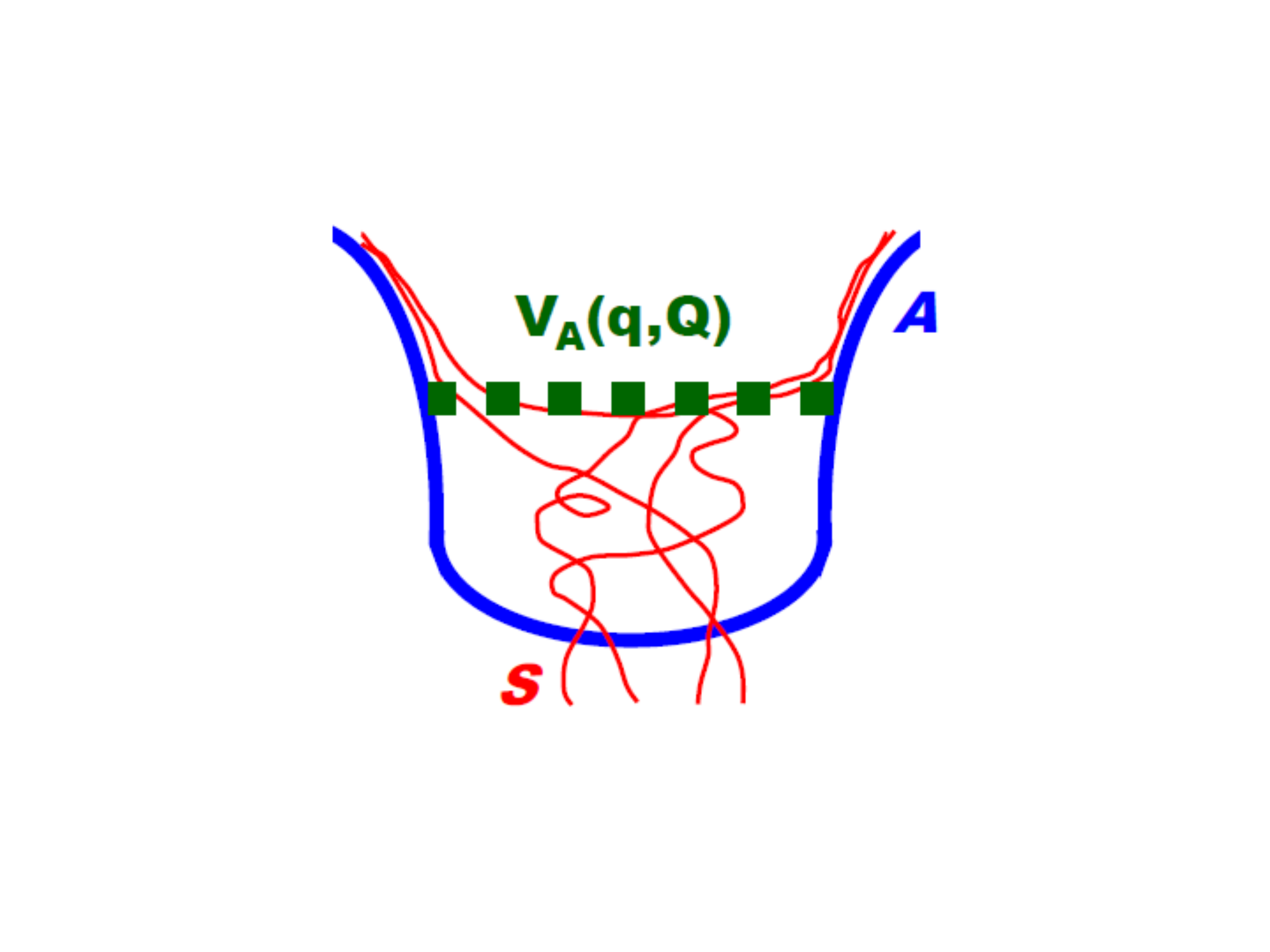}
 \vspace{-2.0cm}
\caption{Typical paths for a microscopic system ${\cal S}$ (shown in red) and a "macroscopic system" ${\cal A}$, shown in blue; during the time the interaction $V_A(q,Q)$ is switched on, it gradually correlates the paths of ${\cal S}$ with those of ${\cal A}$.   }
 \label{fig:NJP-F15}
\end{figure}

In any case it is clear that, once the measurement interaction has finished, the path bunching mechanism will operate simultaneously on the paths of the combined system ${\cal S} + {\cal A}$; indeed the paths of ${\cal S}$ will be "slaved" to those of ${\cal A}$, provided the interaction $V_M(q,Q)$ has done its job of correlating the paths if the 2 systems. Now, since the the dynamics of $Q$ is essentially classical - its mass exceeds the "critical mass" above which its dynamics becomes classical -  the slaving mechanism then implies that "state superpositions" of different positions $q,q'$ for ${\cal S}$ will no longer be possible either, after the measurement. The general behaviour is shown in Fig. \ref{fig:NJP-F15}.


Actually the analysis of the dynamics of this measurement in CWL theory is not so simple - as  ${\cal S}$ begins to "feel" the path bunching mechanism via its interaction with ${\cal A}$, various transients occur in its dynamics (ie., in the behaviour of its paths). The details of this are
rather lengthy, and will be given elsewhere \cite{GR-Psi-1}.

Finally, we can ask whether one can define something analogous to a "wave-function" in CWL theory. The problem here is that any such wave-function must satisfy an infinite hierarchy of integro-differential equations, involving a succession of higher "self-consistent" potentials (which resemble the "superpotentials" introduced by Chandrasekhar and Lebowitz \cite{chandra62} in another context). Although these can be useful in practical application of the theory \cite{GR-Psi-1}, they are not essential to its basic formulation. For all the reasons given in this paper, it is preferable to think of paths and "path space" as non-local objects which give a truer picture of the underlying physics.


\section{Concluding Remarks}
 \label{sec:fin}


The principal object of this paper has been to give a relatively informal discussion of the ideas leading to the CWL theory. We see that the main punchline here is that it is possible to devise an internally consistent theory in which the breakdown of QM is caused by gravitational correlations. The theory is also consistent with Einstein GR at energies which certainly cover lab scales - indeed, we expect it to be consistent with GR for massive bodies, up the energies currently tested in astrophysics. It will however not be consistent with QM, in any experiments designed to look for real-space QM correlations and interference on large masses. In this sense the CWL theory provides an objective distinction between classical macroscopic and QM microscopic processes. We also re-emphasize that the rapid crossover then predicted between quantum and classical dynamics, caused by "path bunching", is {\it not} a decoherence process.

At this point it is useful to re-examine the 2 thought experiments considered at the beginning of this paper (section \ref{sec:thoughtExp}). Does the CWL theory solve the paradoxes which we argued must exist, when including gravity in a 2-path interference or an EPR experiment?  We argue that it does - for if path bunching occurs, the suppression of any interference effects for separated paths means that all metrics contributing now to the path integral are now effectively the same. Thus we no longer have a superposition of different spacetime metrics, except for small masses.

We see that we have arrived at a theory in which both matter and the spacetime metric behave quantum-mechanically for small masses; but once we get to larger masses, these superpositions no longer play a role in the theory, and it can be treated classically. Clearly the fact that one can devise such a theory, which is internally consistent, is interesting in itself, purely as a theoretical result. It will be certainly more interesting if its predictions turn out to be correct in experiments.

Clearly to discuss experiments it is necessary to look in much more detail at how the CWL theory works for an extended massive body. This task is sufficiently important that it is discussed in a separate paper \cite{GR-Psi-1}. As noted already, the results turn out to be depend in a rather crucial way on the detailed structure of the body - this will actually make it easier to test the theory.

Finally, we can ask what other things need to be done. Quite apart from a more detailed discussion of the theory itself, the most pressing current questions are:

(i) How does one deal with situations in which the spacetime curvature is strong? If the spacetime is static, then the obvious tactic is to adapt to CWL theory the existing methods used to treat quantum fields theory on background curved spacetime \cite{wald-QFT}; but if it is strongly time-dependent things are not at all obvious. Important things to clarify include the calculation of vacuum energy in the CWL theory (a calculation which is contentious even in conventional theory \cite{wald04}), and the physics near an event horizon.

(ii) What sort of interesting phenomena do we expect to see in the ultra-relativistic regime in this theory? The behaviour of photons already turns out to be very interesting; but what happens to massive particles at very high energies, when their relativistic mass approaches the "path bunching" crossover regime; and what are the implications for physics in the early universe?

(iii) What is the role of torsion in this theory? The thought experiments discussed in section \ref{sec:CWT-arg} give no role to torsion; but the formal structure of CWL theory certainly allows for it as a possible ingredient.

(iv) What role does environmental decoherence play in the physics? This question is highly relevant to experiments \cite{GR-Psi-1}. We may distinguish four kinds of decoherence here. There is standard environmental decoherence from a variety of oscillator bath \cite{weiss} and spin bath \cite{RPP00} sources; there is decoherence from gravitons \cite{blencowe}, which although it is gravitational in origin, is also just another conventional oscillator bath decoherence mechanism; there is "3rd-party decoherence" \cite{stamp06}, whereby a system is entangled with some environment via a 3rd system, even though the system and environment never interact (the mechanisms recently discussed by Zych et al. \cite{pikowski} and Gooding and Unruh \cite{cisco14} are examples of this); and then there is genuine intrinsic decoherence, lying outside conventional QM, of the kind discussed by Penrose, 't Hooft, and others. To sort all of these out in experiments (and distinguish them from the CWL mechanism discussed here, which is not a decoherence mechanism at all) will be an interesting experimental challenge!


\section{Acknowledgements}
 \label{sec:ackn}


This paper and the ideas in it have benefited greatly from discussions with Harvey Brown, Roger Penrose, Bill Unruh, and Bob Wald. I would also like to thank Cisco Gooding, Friedemann Queisser and Gordon Semenoff for their discussions and collaborations on related work. The work was supported by NSERC of Canada, and by the Templeton foundation, grant No. 36838.




\begin{thebibliography}{99}
\bibitem{repeatP}      In many cases the formal exclusion of 2 paths which overlap is unnecessary, because these paths are of measure zero. However one needs to be careful in restricted geometries, because the contribution from 2 paths that do overlap is divergent, and if paths are forced through a constriction, the extra contribtion may not be negligible.
\end{thebibliography}
\end{document}